\documentclass{article}

\usepackage[a4paper,margin=1in]{geometry}
\usepackage{amsmath}
\usepackage{amssymb}
\usepackage{graphicx}
\usepackage{float}
\usepackage{xcolor}
\usepackage{enumitem}
\usepackage{authblk}
\usepackage{url}
\usepackage{tabularx}
\usepackage{booktabs}
\usepackage{array}
\usepackage{tikz}
\usetikzlibrary{arrows.meta,positioning,fit,backgrounds,calc}
\usepackage[colorlinks=true,linkcolor=blue,citecolor=blue,urlcolor=blue]{hyperref}

\title{Scattering Amplitudes as Programs:\\Self-Evolving Search for Theory and Event Generation}
\author[1]{Yi Gu}
\author[1,2]{Sven Krippendorf}
\affil[1]{Department of Applied Mathematics and Theoretical Physics, University of Cambridge}
\affil[2]{Cavendish Laboratory, University of Cambridge}
\date{\today}

\begin{document}

\maketitle

\begin{abstract}
By viewing scattering amplitudes as computer programs, we connect two goals: exposing useful analytic structure and constructing efficient numerical evaluators for collider phenomenology. Equivalent programs can differ sharply in multiplicity scaling, arithmetic complexity, cancellation, and runtime. Amplitude calculation therefore defines a structured search problem over analytic representations, recursive algorithms, colour and helicity organisation, and reuse of intermediate objects.

We embed repository-scale coding agents inside an external generate--evaluate--select loop with frozen evaluators and objectives, and study three optimisation targets. A scaling search moves from BCFW recursion to a specialised fixed-\(k\) split-helicity transfer algorithm, reaching an \(805\times\) geometric-mean speed-up on the scored grid. A structural search reorganises NMHV terms into R-invariant-style cells and glued supercells, reducing inter-term cancellation. Across twenty QCD and electroweak processes, an exact-operation search reduces counted arithmetic by \(47.7\times\) and gives a \(5.9\times\) post-hoc Python runtime improvement. In matched pure-gluon component tests from \(n=4\) to \(n=6\), the evolved engine's speed advantage over the tested Sherpa-Comix exact-sum call grows from \(10\times\) to \(277\times\), while its gap to process-specialised MadGraph5\_aMC@NLO Fortran compiled at \texttt{-O3} narrows from \(77.7\times\) to \(13.7\times\).

The searches move between mathematical representations and combine recursion, symmetry, basis reduction, dynamic programming, and shared computation into hybrid amplitude programs. They provide initial evidence that parts of amplitude optimisation can be made systematic through self-evolving program search, while leaving native generator integration and end-to-end event throughput as future tests. The generator comparison is an isolated exact matrix-element call, not a modification of MadGraph or Sherpa.
\end{abstract}

\clearpage

\tableofcontents

\section{Introduction}

Precision collider physics repeatedly evaluates scattering amplitudes inside event generators, fixed-order calculations, matching and merging, and detector-level simulation~\cite{Alwall:2014hca,Gleisberg:2008ta,Campbell:2022qmc}. At HL-LHC scale this is not a marginal software cost. The conservative and aggressive Run-4 projections assign \(17\%\) and \(20\%\), respectively, of the ATLAS CPU budget to event generation (Fig.~3(a,b) of Ref.~\cite{ATLAS:2022hllhcroadmap}), while a CMS Phase-2 estimate projects approximately \(79\) billion simulated events per year in Run~4 and \(100\) billion per year in Run~5 (Table~1 on p.~4 of Ref.~\cite{CMS:2022phase2computing}). Limited Monte-Carlo statistics already constrains precision measurements~\cite{Valassi:2021eik}.

Vector-boson-plus-jets production makes the pressure concrete. An ATLAS study estimates that samples corresponding to \(3\,\mathrm{ab}^{-1}\) require roughly \(330\) billion \(V+\)jets events and \(3.8\) million HS06-years even with an improved Sherpa~2.2.11 setup~\cite{ATLAS:2022hsp}. In representative merged-NLO samples, phase-space generation and tree-level matrix elements account for more than half of the runtime~\cite{Bothmann:2022qkz}. The practical target is therefore not merely to obtain a formula, but to evaluate the relevant maps scalably inside production chains.

MadGraph and Sherpa make such calculations systematic, from process generation through matrix elements and phase-space integration to shower interfaces and event production~\cite{Alwall:2014hca,Gleisberg:2008ta}. Their success makes generator kernels natural reference points, while recent Sherpa improvements and data-parallel matrix-element approaches show that algorithmic and implementation changes can buy large factors~\cite{Bothmann:2022qkz,Bothmann:2023pepper,Hagebock:2025mg5cudacpp}. This paper studies the interface between amplitude structure and executable efficiency.

The theoretical history of amplitudes can itself be read as a sequence of better computational concepts: spinor-helicity variables~\cite{Kleiss:1985yh,Dixon:1996wi}, colour ordering~\cite{Mangano:1987xk,Dixon:1996wi}, reusable Berends--Giele currents~\cite{Berends:1987me}, on-shell recursion~\cite{Britto:2004ap,Britto:2005fq}, MHV-vertex expansions, unitarity~\cite{Bern:1994zx}, colour--kinematics relations, and geometric representations~\cite{Arkani-Hamed:2013jha}. These ideas did not only shorten derivations; they changed the executable route. Equivalent expressions can differ substantially in scaling, cancellation, reuse, and suitability for generated code. A mathematical representation is therefore also a computational design principle.

Learned symbolic methods provide one nearby precedent. Transformers have exposed regularities in the planar \(\mathcal N=4\) bootstrap~\cite{Cai:2024transformingBootstrap,Cai:2025recurrentFeatures}. Dersy, Schwartz and Zhang treat polylogarithm simplification as reinforcement learning and sequence translation~\cite{Dersy:2022polylog}; Cheung, Dersy and Schwartz train a specialised transformer to simplify spinor-helicity expressions and use learned embeddings to select mutually simplifying term subsets~\cite{Cheung:2025simplicity}. These systems learn a simplification map from examples. Here, the desired final program need not be known: executable candidates are instead judged by correctness and explicit physical or computational objectives.

Agentic scientific systems broaden this setting. Coding agents have modified CLASS under compilation, reference-observable and cosmological-data checks~\cite{Mudur:2025cosmomodel}; DiscoverPhysics couples experiment design to executable inference in simulated worlds~\cite{Wiemann:2026discoverphysics}. MadEvolve is the closest architectural comparison, evolving populations of cosmological algorithms under numerical objectives while optimising continuous parameters in a separate inner loop~\cite{Li:2026madevolve}. LLM-assisted amplitude and wave-scattering studies provide complementary examples of formula discovery~\cite{Guevara:2026singleminus,ArkaniHamed:2026hydrotope}.

Within perturbative quantum field theory, program search has also optimised integration-by-parts reduction. FunSearch-style priority functions reduce seeding integrals in difficult examples, while typed genetic programming recovers and sometimes improves established heuristics~\cite{Song:2025funsearchIBP,vonHippel:2025funsearchIBP}. We move the same principle to repository-scale amplitude evaluators, study multiplicity scaling directly, and target matrix-element kernels whose acceleration can ultimately make event generators faster.

Recent work has also combined reinforcement learning, evolutionary strategies and coding agents to rediscover and vary tube-seeding strategies for integration-by-parts reduction~\cite{Berman:2026tubeseeding}. This provides a further perturbative-QFT precedent for evaluator-guided algorithm search, while the present work searches complete executable amplitude programs and follows representation changes across both formal and generator-relevant objectives.

A repository-scale coding agent can inspect interacting modules, edit code, execute tests and profilers, diagnose failures, and repair a proposal before submission. Embedded in an external generate--evaluate--select loop, it becomes the mutation operator of a \emph{self-evolving} system: the population of executable programs evolves under external evaluation, while the model weights, evaluator, and objective remain fixed within each reported run. The coding agent alone is not the experiment; the frozen ruler and population dynamics provide the selection pressure.

This combination admits changes of representation rather than only parameter tuning. The searches reported below move from BCFW recursion through split-helicity formulae to a transfer dynamic program; from CSW terms to R-invariant-style cells and glued supercells; and from process-by-process routines to hybrid engines combining recursive currents, colour-basis reduction, helicity and parity relations, chiral algebra, and shared contractions. Such hybrids are natural because no single named formalism simultaneously optimises scaling, exact arithmetic, cancellation, runtime, memory, and portability.

Domain expertise remains essential. It fixes physical regimes, trusted references, adversarial gates, meaningful objectives, and the evidence required before a candidate is interpreted as an amplitude result or a production optimisation. The agent supplies breadth and iteration speed across a combinatorial implementation space. The division of labour is particularly valuable for amplitudes, where candidate programs are both executable artefacts and inspectable mathematical objects.

To our knowledge, this is the first use of repository-scale coding agents as mutation operators in an externally evaluated evolutionary search over complete executable scattering-amplitude programs. The paper makes three contributions: a common executable formalism for analytic and phenomenological optimisation; controlled examples in which search traverses recognisably different amplitude representations; and initial evidence for sizeable gains in generator-relevant matrix-element kernels. These are tree-level benchmark studies, not an end-to-end event-generation result.

The frozen artefacts, full lineages, evaluators, prompts, databases and reproduction scripts are archived in the \href{https://github.com/YiGu310/scattering-amplitudes-program-search/tree/ab8c8682713ff8592bb55e456229619bff7cfd12}{versioned companion repository (commit \texttt{ab8c868})}~\cite{GuKrippendorf:2026repository}.

The rest of the paper is structured as follows. Section~\ref{sec:amplitudes-as-programs} formulates scattering-amplitude calculations as executable maps and introduces the correctness gates and optimisation objectives. Section~\ref{sec:search-design} describes the self-evolving generate--evaluate--select system and its experimental protocols. Section~\ref{sec:results} presents three searches probing distinct parts of the amplitude-program space: multiplicity scaling, local analytic cancellation, and generator-relevant arithmetic complexity. Section~\ref{sec:discussion} interprets the resulting transitions and hybrid programs, discusses their phenomenological implications, and outlines the path toward native event-generator integration.

\section{Scattering Amplitudes as a Program-Search Problem}
\label{sec:amplitudes-as-programs}

A scattering amplitude is a physical quantity fixed by the theory, external states and kinematics, but its calculation can be represented as an executable program. In a formal calculation that program may be expressed as a closed formula, a recursion relation, a sum over diagrams, or a decomposition into basis functions. In a phenomenological workflow it is literally a piece of software: given a process specification and a phase-space point, a matrix-element routine returns a complex amplitude, a squared matrix element, or an event weight. We regard different amplitude methods as different programs computing the same target quantity and ask how such programs should be represented, checked, compared and searched over.

This section sets up that viewpoint. Section~\ref{subsec:amplitudes-executable-maps} defines the amplitude computation as a map from physics input to numerical output. Section~\ref{subsec:amplitudes-benchmark-history} explains why tree-level gluon amplitudes are a natural first benchmark and why the history of amplitude methods can be read as a history of program improvements. Section~\ref{subsec:metrics-objectives} defines the admissibility gates, metrics, scaling fits, and optimisation objectives. Section~\ref{subsec:evaluation-and-search} then explains how these ingredients are used to evaluate searched programs relative to existing solutions.

\subsection{Amplitudes as executable maps}
\label{subsec:amplitudes-executable-maps}

At the most abstract level, an amplitude program implements a map
\begin{equation}
    \mathrm{output}=P(\mathrm{input}) .
    \label{eq:program-map}
\end{equation}
The input contains the physical problem, while the program specifies a computational route to the answer. For a fixed benchmark instance we write schematically
\begin{equation}
    x=(\mathcal T,\mathcal Q,n,h,\Phi,\eta),
\end{equation}
where $\mathcal T$ denotes the theory or model, $\mathcal Q$ the process specification including particle species and flavour labels, $n$ the external multiplicity, $h$ the helicity assignment, $\Phi$ a phase-space point, and $\eta$ collects masses, couplings, colour representation data, numerical settings, and the requested perturbative order or approximation. A candidate program $P$ maps this data to one of several possible outputs:
\begin{equation}
    P(x)\in \{A_n,\,\mathcal M_n,\,\Sigma_n,\,\overline{|\mathcal M_n|^2},\,w_{\rm event}\} .
\end{equation}
Here $A_n$ is a colour-ordered partial amplitude, $\mathcal M_n$ a colour-dressed amplitude, $\Sigma_n$ its unaveraged colour--helicity sum, $\overline{|\mathcal M_n|^2}$ the initial-state-averaged quantity defined below, and $w_{\rm event}$ the quantity eventually consumed by an event-generation workflow.

The distinction between physics input and program choice is important. The same physical map can be evaluated by Feynman diagrams, colour-ordered formulae, Berends--Giele currents, Britto--Cachazo--Feng--Witten (BCFW) recursion, Cachazo--Svr\v{c}ek--Witten (CSW) diagrams, generated helicity code, or a production matrix-element generator. These are not different physical theories; they are different programs for the same theory. Their outputs agree when they are correct, but their operation counts, intermediate building blocks, memory use, numerical stability, and scaling can be very different.

For tree-level gluon amplitudes we use the standard colour-ordered spinor-helicity language~\cite{Mangano:1990by,Dixon:1996wi,Dixon:2013uaa,Elvang:2013cua}. In these conventions the colour-dressed amplitude admits the single-trace decomposition
\begin{equation}
\mathcal{M}_n^{\mathrm{tree}}
\!\left(\{p_i,h_i,a_i\}_{i=1}^{n}\right)
=
 \mathcal N_n g^{\,n-2}
\sum_{\sigma\in S_n/Z_n}
\mathrm{Tr}\!\left(T^{a_{\sigma(1)}}\cdots T^{a_{\sigma(n)}}\right)
A_n^{\mathrm{tree}}\!\left(
p_{\sigma(1)}^{\,h_{\sigma(1)}},\ldots,
p_{\sigma(n)}^{\,h_{\sigma(n)}}\right),
\label{eq:single-trace-decomposition}
\end{equation}
where the sum is over non-cyclic permutations. Here the permutation acts on complete external-leg data: momentum, helicity and colour label move together. We use \(T^a=\lambda^a/2\), so that \(\mathrm{Tr}(T^aT^b)=\delta^{ab}/2\). Convention-dependent normalisation factors and the conversions used by the executable benchmarks are documented in Appendix~\ref{app:processes}.

The symmetries of partial amplitudes, colour organisation, and helicity sums are already algorithmic decisions. We distinguish the full sum from its incoming-state average:
\begin{align}
\Sigma_n(\Phi)
&=
\sum_{\{h\}}\sum_{\{a\}}
\left|\mathcal M_n\!\left(\{p_i,h_i,a_i\}\right)\right|^2,
\nonumber\\
\overline{|\mathcal M_n|^2}(\Phi)
&=
\frac{\Sigma_n(\Phi)}
{N_{\mathrm{spin,in}}N_{\mathrm{colour,in}}}.
\label{eq:squared-matrix-element}
\end{align}
The first quantity is the unaveraged colour- and helicity-summed squared matrix element. The second additionally averages over the spin and colour states of the incoming particles and is the convention used in the tree-complexity benchmark. For two incoming gluons in four dimensions, \(N_{\mathrm{spin,in}}=2^2\) and \(N_{\mathrm{colour,in}}=(N_c^2-1)^2\). Final-state phase-space symmetry conventions used by individual benchmark engines are stated separately in Section~\ref{subsec:results-generators} and Appendix~\ref{app:processes}. The partial-amplitude benchmarks use fixed ordering and helicity and include neither sums nor averages.

Perturbative order gives another example of the same separation between physics target and program implementation. A generic amplitude may be written as
\begin{equation}
    \mathcal M
    =
    \mathcal M^{(0)}
    +
    \mathcal M^{(1)}
    +
    \mathcal M^{(2)}+\cdots ,
    \label{eq:perturbative-expansion}
\end{equation}
where \(\mathcal M^{(\ell)}\) is the \(\ell\)-loop amplitude and coupling powers are understood. The benchmark developed here starts with tree-level building blocks. Loop-level programs may additionally include integral reduction, master-integral evaluation, subtraction terms and stability fallbacks~\cite{Bern:1994zx,Laporta:2000dsw,Catani:1996vz}.
The tree-level amplitude \(\mathcal M^{(0)}\) supplies the leading-order contribution, while \(\mathcal M^{(1)}\) and \(\mathcal M^{(2)}\) enter the virtual parts of NLO and NNLO predictions. The corresponding fixed-order cross sections additionally contain real-emission, interference and subtraction contributions.

Finally, we distinguish the physics input from representation and implementation choices. These choices include the recursion or decomposition used, the choice of basis, caching and reuse of intermediate objects, numerical precision, expression simplification, data layout, compiler settings, and hardware. Some of these choices are device independent, in the sense that they change the mathematical or algorithmic representation. Others are device dependent, in the sense that they affect a concrete implementation on a particular architecture. The metric definitions below keep these categories separate.

\subsection{Why tree-level gluons form a benchmark}
\label{subsec:amplitudes-benchmark-history}

Tree-level gluon amplitudes are a useful first target because they are simple enough to admit transparent seed programs, but rich enough to contain much of the algorithmic structure that makes amplitudes interesting. They have compact special cases, nontrivial general helicity sectors, a well-developed set of analytic and recursive methods, and trusted implementations in event generators. They also make the central point of this paper especially vivid: progress in amplitudes often looks like the discovery of a better program.

The Parke--Taylor formula showed that an all-multiplicity MHV amplitude can collapse to a one-line expression, revealing a simplicity that is obscured by a diagrammatic expansion~\cite{Parke:1986gb}. Berends--Giele recursion reorganised tree amplitudes into reusable off-shell currents, replacing a direct enumeration of diagrams by a dynamic-programming calculation~\cite{Berends:1987me}. The twistor-string perspective and the CSW construction recast MHV amplitudes as vertices for more general tree amplitudes~\cite{Witten:2003nn,Cachazo:2004kj}. BCFW recursion then expressed tree amplitudes in terms of lower-point on-shell amplitudes, turning factorisation and analyticity into an executable divide-and-conquer algorithm~\cite{Britto:2004ap,Britto:2005fq}. Later developments, including colour-kinematics relations and geometric structures such as the amplituhedron, further illustrate that changing representation can change the apparent complexity of the calculation~\cite{Bern:2008qj,Arkani-Hamed:2013jha}.

For our purposes this history is not merely background. It provides a set of known transformations that a program-search system ought to be able to value, and ideally to recover in controlled settings. The transformations include compression of expressions, replacement of diagrams by reusable building blocks, introduction of recursive subproblems, cancellation of spurious intermediate structure, and improved scaling with multiplicity. These are all naturally expressible as changes in executable code.

The first two benchmarks focus on controlled colour-ordered partial amplitudes $A_n$: one optimises multiplicity scaling and one optimises the rational term structure. This isolates kinematics from colour and helicity sums. The more phenomenological benchmark instead returns the incoming-state-averaged \(\overline{|\mathcal M_n|^2}\), with the separately stated final-state symmetry convention, across a twenty-process suite and minimises exactly counted arithmetic operations. Its references are MadGraph calculations or project engines previously validated against MadGraph; it is not a direct benchmark of native MadGraph or Sherpa kernels.

The seed programs should be correct but not over-optimised. For partial amplitudes, compact Python implementations of Parke--Taylor special cases, Berends--Giele recursion, CSW-like constructions, and BCFW recursion provide natural starting points. For the squared-matrix-element suite, a transparent Python implementation is useful because its operation graph can be instrumented exactly and its outputs checked against independently validated references. Python makes the search space inspectable, mutable and easy to instrument; a successful searched idea can later be translated into a production code-generation framework.

\subsection{Metrics and optimisation objectives}
\label{subsec:metrics-objectives}

The benchmark separates correctness from optimisation. Correctness is not one objective among many; it is the ticket of admission. After that gate has been passed, different families of metrics can be used either for selection or only for reporting. This distinction is important because the selected score in a given run may use only one subset of the available measurements.

\paragraph{Hard admissibility gates.}
A candidate program first has to be admissible. In the present context this means that it parses or compiles, satisfies the benchmark input/output contract, executes on the benchmark inputs, returns finite numerical values, and passes the required correctness tests. The three frozen evaluators use the following complex relative discrepancy for a reference program \(P_{\rm ref}\) and validation set \(\mathcal V\):
\begin{equation}
    \epsilon_{\rm rel}(P)
    =
    \max_{x\in\mathcal V}
    \frac{|P(x)-P_{\rm ref}(x)|}
         {\max\{|P_{\rm ref}(x)|,10^{-300}\}} .
    \label{eq:relative-error}
\end{equation}
The modulus is taken after complex subtraction; the tiny denominator guard only prevents division by zero and is not an absolute-error scale. The BCFW-scaling scattered-helicity gate uses tolerance \(10^{-6}\), with a ladder from \(10^{-6}\) to \(3\times10^{-3}\) only on its deepest scored cells. The structural run uses \(10^{-6}\) on five gate points per configuration, and the tree-complexity run uses \(10^{-6}\) on four fresh points for each of twenty processes. Non-finite scaling outputs are rejected explicitly; in the other two experiments a non-finite ratio fails the strict comparison or a dedicated finite-numerator gate. Candidates that fail any gate receive score zero.

\paragraph{Device-independent algorithmic metrics.}
For admissible programs, one can measure properties of the algorithmic representation rather than of a particular machine. The main device-independent quantities are operation counts, counts of computational building blocks, and program or expression length:
\begin{equation}
    \mathbf m_{\rm alg}(P;n)
    =
    \big(N_{\rm op}(P;n),\,N_{\rm block}(P;n),\,S_{\rm prog}(P)\big) .
    \label{eq:algorithmic-metrics}
\end{equation}
Here $N_{\rm op}$ denotes a fixed operation-count proxy, for example the number of arithmetic operations, complex operations, spinor products, current combinations, or expression-tree operations, depending on the benchmark. $N_{\rm block}$ is used only when a natural unit of computation has been fixed, such as Berends--Giele currents, BCFW terms, CSW diagrams, colour structures, propagators, or cached subcurrents. $S_{\rm prog}$ measures program length, such as lines of code, abstract-syntax-tree size, number of generated expressions, or size of a symbolic representation. These metrics need not rank programs in the same way: a compact formula may be expensive to evaluate, while a longer recursive implementation may scale better.

\paragraph{Device-dependent performance metrics.}
A second family of metrics measures the concrete implementation on a specified hardware and software stack,
\begin{equation}
    \mathbf m_{\rm dev}(P;n,H)
    =
    \big(T(P;n,H),\,M_{\rm peak}(P;n,H),\,U(P;n,H)\big) ,
    \label{eq:device-metrics}
\end{equation}
where $H$ denotes the environment. $T$ may be wall-clock time, median per-point evaluation time, throughput, or total benchmark time. $M_{\rm peak}$ records peak memory use or a suitable memory-allocation proxy. $U$ denotes resource utilisation when that is available. These quantities matter for phenomenology because production workflows run on real hardware, but they must be reported together with the hardware, language, compiler, parallelisation, and data layout.
Here an environment \(H\) means the complete timed stack: processor and memory, operating system, language runtime, compiler and flags, numerical libraries, thread settings and data layout. Holding \(H\) fixed makes comparisons within one benchmark meaningful; changing it can alter absolute times and sometimes their ranking.

\paragraph{Selection scores and reported metrics.}
A given search run uses a decisive score built from a chosen subset of the metrics above, after the hard gate has been passed. The scaling run assigns \(75\%\) of its score to fitted timing exponents and \(25\%\) to geometric-mean speed. The structural run uses a summation-conditioning proxy defined in Equation~\eqref{eq:structural-score}. The twenty-process run selects only on exactly counted real arithmetic operations, with runtime withheld from the agent and reported afterwards. These are experimental protocols, not a universal loss function.

\paragraph{Scaling behaviour.}
Many amplitude algorithms are valuable because they change how the cost grows with multiplicity. For the first experiment we fit \(\log T=a+p\log n\) by least squares on the stated multiplicity window. The fitted slope \(p\) is an effective finite-window exponent, not automatically an asymptotic complexity. We therefore state the fit window and, in Section~\ref{subsec:results-gluon}, report a separate source-level and timing audit extending the champion to \(n=256\). Selection uses the best of two or three repeats, appropriate for one-sided timing interference; the reported scaling figure instead uses medians over three phase-space points and five repeats per point, with interquartile ranges.

\paragraph{Pareto viewpoint.}
The metric choices are illustrative rather than unique. A program \(P\) Pareto-dominates \(P'\) if it is no worse on every reported metric and strictly better on at least one. A non-dominated candidate can remain scientifically interesting even when it does not maximise the scalar selection score---for example, it may trade a slightly larger operation count for lower memory or better high-multiplicity scaling. The present runs select with one declared scalar score while retaining secondary measurements.

\subsection{Evaluation and search targets}
\label{subsec:evaluation-and-search}

The benchmark separates two questions: does a candidate compute the required quantity, and, if so, how well does it meet the declared objective? A candidate first passes parsing, interface and numerical-correctness gates. Only then is its selection score evaluated. Secondary metrics are retained for later diagnosis and for comparisons with the seed, analytic references and, where appropriate, established tools.

The optimisation and validation samples are disjoint. In the BCFW-scaling run, a \emph{timing cell} is one pair \((k,n)\) of negative-helicity count and external multiplicity; the score aggregates 34 such cells. A separate 19-case gate includes randomly drawn phase-space points and non-contiguous, or ``scattered'', negative-helicity arrangements. In the structural run, a \emph{point pool} is the fixed set of 300 phase-space points used only to measure cancellation for one helicity configuration; correctness is tested elsewhere, including at \(n=8\). In the tree-complexity run, exact operation counts are evaluated at one fixed point generated with RAMBO (Random Momenta Beautifully Organised)~\cite{Kleiss:1985gy} because the counted program is forbidden to branch on numerical data; every candidate is additionally checked at four new points for each process. These gates test transfer across phase-space points and selected helicity arrangements, but not to untested process families or arbitrarily large multiplicity.

Different targets favour different representations: runtime includes machine-level effects, exact operation count probes the arithmetic graph, scaling rewards flatter multiplicity dependence, and the cancellation score favours more local cancellation near spurious boundaries. The intended theoretical or phenomenological use therefore fixes the appropriate target.

This completes the physics-side setup of the paper. We have identified the executable objects to be searched over, the historical amplitude structures that motivate the benchmark, the hard correctness gates, the algorithmic and device-dependent metrics, and the optimisation objectives used to compare programs. The next section turns to the algorithmic question: how to search this program space in practice.

\section{Self-Evolving Program Search}
\label{sec:search-design}

\begin{figure}[t]
\centering
\begin{tikzpicture}[
  font=\small,
  mainbox/.style={
    draw=black!70, fill=white, rounded corners=2.5pt, align=center,
    minimum height=1.05cm, minimum width=2.55cm, inner sep=5pt
  },
  auxbox/.style={
    draw=black!55, fill=white, rounded corners=2pt, align=center,
    minimum height=0.70cm, inner sep=4pt, font=\footnotesize
  },
  envbox/.style={
    draw, dashed, rounded corners=5pt, inner sep=10pt
  },
  envtitle/.style={font=\footnotesize\scshape, anchor=south west, inner sep=2pt},
  badge/.style={
    circle, fill=black!75, text=white, font=\bfseries\footnotesize,
    inner sep=0.8pt, minimum size=0.42cm
  },
  islandbox/.style={
    draw=black!50, dashed, fill=white, rounded corners=1.5pt,
    minimum width=1.30cm, minimum height=0.92cm, inner sep=2pt
  },
  flow/.style={-{Latex[length=2.6mm]}, line width=0.9pt, draw=black!75},
  auxflow/.style={-{Latex[length=1.9mm]}, line width=0.5pt, draw=black!55, dashed},
  flowlabel/.style={font=\footnotesize, align=center},
]
\newcommand{\progcard}[2]{%
  \begin{scope}[shift={#1}]
    \draw[#2, line width=0.5pt, rounded corners=0.8pt, fill=white]
      (0,0) rectangle (0.30,0.40);
    \draw[#2, line width=0.4pt] (0.055,0.29) -- ++(0.19,0);
    \draw[#2, line width=0.4pt] (0.055,0.21) -- ++(0.19,0);
    \draw[#2, line width=0.4pt] (0.055,0.13) -- ++(0.11,0);
  \end{scope}
}

\node[mainbox, minimum width=3.05cm] (agent) at (0,0)
  {Repository-scale agent\\ edit--test--repair};

\node[auxbox, above=0.55cm of agent] (context)
  {Repository context $\Omega_i$\\ prompts, tools, budget};

\node[auxbox, below=0.55cm of agent, xshift=0.45cm] (localtests)
  {Local tests visible\\ to the agent};

\node[mainbox, right=0.95cm of agent, minimum width=2.70cm] (candidate)
  {Submitted candidate\\ $P^{(a)}_{i+1}$};

\draw[auxflow] (context) -- (agent);
\coordinate (ltL) at ($(localtests.north)+(-0.30,0)$);
\coordinate (ltR) at ($(localtests.north)+(+0.30,0)$);
\draw[auxflow, solid] (ltL |- agent.south) -- (ltL);
\draw[auxflow, solid] (ltR) -- (ltR |- agent.south);
\draw[flow] (agent) -- (candidate);

\node[mainbox, right=2.05cm of candidate, minimum width=3.30cm] (evaluator)
  {Frozen evaluator\\ {\footnotesize correctness gates + objective}};

\node[mainbox, below=0.80cm of evaluator, minimum width=3.30cm] (score)
  {Score and feedback\\ $s_i(P),\ \mathcal E_i(P)$};

\draw[flow] (evaluator) -- node[flowlabel, right=1.5pt] {$\mathrm{Eval}$} (score);

\node[font=\small, anchor=north] (poptitle) at ($(score.south)+(0,-1.55)$)
  {Candidate population};

\node[islandbox, below=0.14cm of poptitle.south, xshift=-0.98cm] (isl1) {};
\node[islandbox, below=0.14cm of poptitle.south, xshift=+0.98cm] (isl2) {};
\node at ($(isl1.east)!0.5!(isl2.west)$) {$\cdots$};
\node[font=\tiny, below=0.02cm of isl1] (isl1lab) {$\mathcal I_{i,1}$};
\node[font=\tiny, below=0.02cm of isl2] (isl2lab) {$\mathcal I_{i,K_i}$};

\progcard{($(isl1.center)+(-0.44,-0.28)$)}{black!55}
\progcard{($(isl1.center)+(-0.11,-0.22)$)}{black!55}
\progcard{($(isl1.center)+(0.22,-0.16)$)}{black!55}
\progcard{($(isl2.center)+(-0.31,-0.25)$)}{black!55}
\progcard{($(isl2.center)+(0.06,-0.18)$)}{black!55}

\node[font=\footnotesize, anchor=north]
  (popmath) at ($(isl1lab.south)!0.5!(isl2lab.south)+(0,-0.06)$)
  {$\mathcal P_i=\bigsqcup_{k}\mathcal I_{i,k}$};

\begin{scope}[on background layer]
\node[envbox, draw=blue!45!black, fill=blue!5,
      fit=(context)(agent)(localtests)(candidate)] (proposalenv) {};
\node[envbox, draw=black!60, fill=black!8,
      fit=(evaluator)(score)] (externalenv) {};
\node[mainbox, inner sep=6pt,
      fit=(poptitle)(isl1)(isl2)(isl1lab)(isl2lab)(popmath)]
      (population) {};
\end{scope}

\coordinate (selx) at ($(localtests.west)+(-0.55,0)$);
\node[mainbox, minimum width=2.90cm] (selection)
  at (selx |- population.center)
  {Parent selection\\ program + feedback};

\node[envtitle, text=blue!35!black]
  at ($(proposalenv.north west)+(0.42,0.02)$)
  (proposaltitle) {Writable proposal environment};
\node[envtitle, anchor=south east, align=right, text=black!75]
  at ($(externalenv.north east)+(-0.08,0.02)$)
  (externaltitle) {Authoritative external\\ environment};

\coordinate (lock) at ($(externaltitle.west)+(-0.22,0.0)$);
\draw[black!70, fill=black!70, rounded corners=0.6pt]
  ($(lock)+(-0.105,-0.095)$) rectangle ($(lock)+(0.105,0.055)$);
\draw[black!70, line width=0.7pt]
  ($(lock)+(-0.058,0.055)$) arc[start angle=180, end angle=0, radius=0.058];

\node[badge] at (proposalenv.north west) {1};
\node[badge] at (externalenv.north west) {2};
\node[badge] at (population.north west) {3};
\node[badge] at (selection.north west) {4};

\draw[flow] (candidate) --
  node[flowlabel, above=3.5pt] (submitlab) {submit}
  (evaluator.west |- candidate);
\progcard{($(candidate.east)!0.5!(evaluator.west|-candidate)+(-0.15,-0.20)$)}{black!60}

\draw[flow] (score.south) --
  node[flowlabel, left=2.5pt, pos=0.55, align=right]
    {archive update\\ $F^{(a)}_i=\big(P_i^{(a)},\mathbf L_i,\mathcal E_i\big)$}
  (population.north);

\draw[flow] (population.west) --
  node[flowlabel, above, pos=0.5] {sample parent}
  node[flowlabel, below, pos=0.5, font=\scriptsize]
    {score-weighted, within an island}
  (selection.east);

\draw[flow] (selection.north) --
  node[flowlabel, right=2.5pt, pos=0.40, align=left]
    {parent + feedback\\ record $F^{(a)}_i$ in prompt}
  (proposalenv.south -| selection.north);

\draw[auxflow] ($(population.south)+(0,-0.50)$) --
  node[flowlabel, at end, below=0.52cm, font=\scriptsize]
    {seed programs (generation 0)}
  (population.south);

\node[flowlabel] (titransition)
  at ($($(selection.north east)!0.5!(population.north west)$)+(0.35,0.72)$)
  {one search step\\
   $T_i:(\mathcal P_i,\mathcal F_i,\Omega_i)\mapsto\mathcal P_{i+1}$};
\draw[black!75, line width=0.6pt, -{Latex[length=1.5mm]}]
  ($(titransition.west)+(-0.24,0.14)$)
  arc[start angle=210, end angle=-60, radius=0.13];

\end{tikzpicture}
\caption{Architecture of the externally evaluated self-evolving program-search system. A repository-scale coding agent modifies a selected executable program inside a writable edit--test--repair environment. The submitted candidate is scored by an authoritative external evaluator containing frozen correctness gates and the declared objective. The resulting score and diagnostics update an island-structured program population and influence subsequent parent selection. The agent cannot modify the references, gates or objective; alternative objectives act as experimental dials selecting different representations of the same physical quantity.}
\label{fig:self-evolving-architecture}
\end{figure}
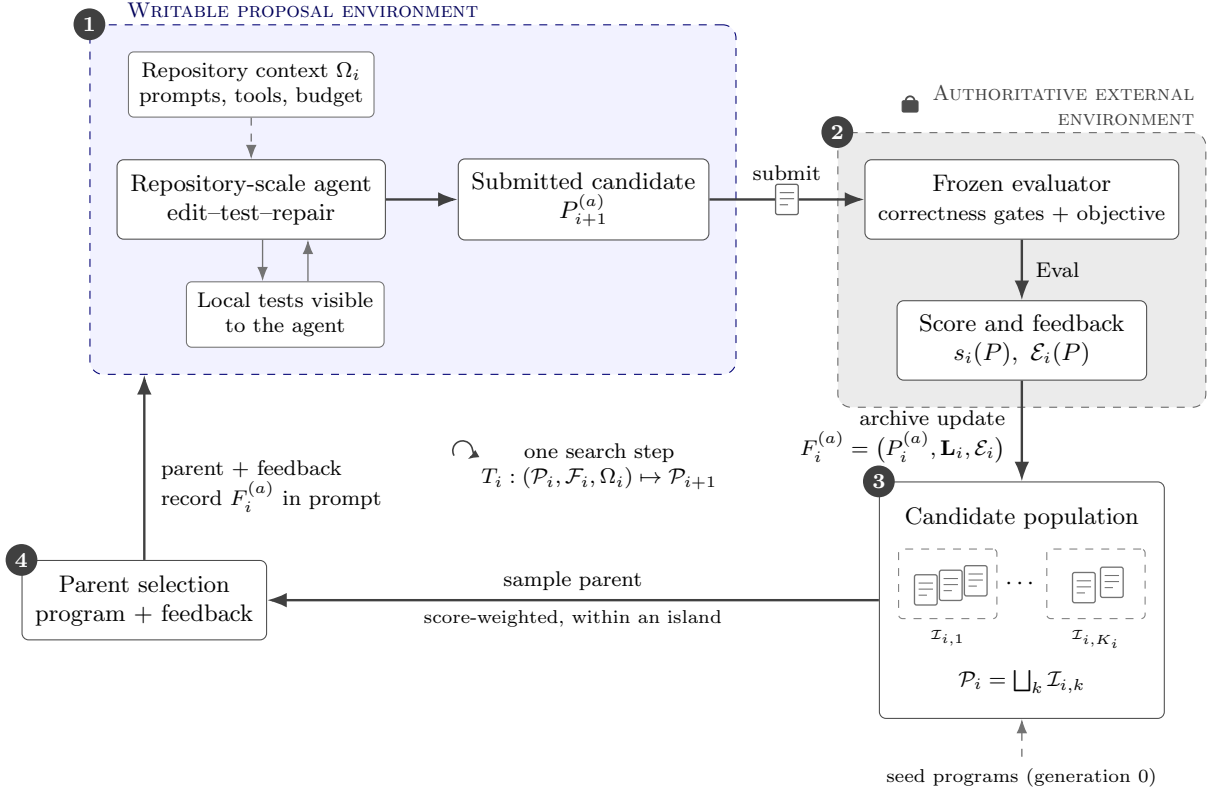

The search maintains executable programs, evaluates them with an external ruler, and uses the resulting records to propose and retain variants. Figure~\ref{fig:self-evolving-architecture} shows the architecture used in the reported searches.

%
%

The evolving objects are executable amplitude programs. A repository-scale coding agent acts as the proposal or mutation operator, but selection is performed by a protected evaluator outside the writable candidate environment. This separation allows the agent to make coordinated changes across mathematical representation and implementation while preventing it from modifying the correctness tests or optimisation objective.

Here \emph{self-evolving} refers to the evolution of the program population and its lineages under repeated external evaluation. The model weights, frozen evaluator, physical target and declared objective remain fixed within each reported run. In the notation below, the population box is \(\mathcal P_i\), measured scores and diagnostics form \(F_i(P)\), the agent plus controller implements \(T_i\), and the frozen selection score is \(s_i(P)\).

This separation between proposal and evaluation also locates the present system within the broader program-search landscape. FunSearch generates short functions under a scalar task score~\cite{RomeraParedes2024FunSearch}; AlphaEvolve permits larger algorithmic edits with population-based retention~\cite{Novikov2025AlphaEvolve}; AdaEvolve adapts search allocation using accumulated progress~\cite{Cemri2026AdaEvolve}; and MadEvolve combines evolutionary search over cosmological algorithms with inner continuous-parameter optimisation~\cite{Li:2026madevolve}. The released searches use ShinkaEvolve's archive, island and generator abstractions~\cite{Lange2025ShinkaEvolve}, with a repository-scale coding-agent session as their proposal operator.

\subsection{Search records and transition operators}
\label{subsec:search-records-transition-operators}

At search step \(i\), let
\begin{equation}
    \mathcal{P}_i = \{P_i^{(1)},\ldots,P_i^{(N_i)}\}
\end{equation}
denote the current population. Each \(P_i^{(a)}\) must implement the benchmark interface, but it may fail to compile, violate the contract, fail numerical validation, or perform poorly. Evaluation produces
\begin{equation}
    \mathrm{Eval}: P_i^{(a)} \longmapsto F_i^{(a)} ,
\end{equation}
with feedback record
\begin{equation}
    F_i^{(a)}
    =
    \Big(
        P_i^{(a)},
        \mathbf{L}_i(P_i^{(a)}),
        \mathcal{E}_i(P_i^{(a)})
    \Big),
    \label{eq:feedback-record}
\end{equation}
where
\begin{equation}
    \mathbf{L}_i(P)
    =
    \big(L_i^1(P),\ldots,L_i^{m_i}(P)\big)
\end{equation}
collects gate and objective components, while \(\mathcal E_i(P)\) contains diagnostics such as errors, residuals, operation counts or timing fits. The complete record can guide an edit even when only one scalar is decisive. A coding-agent edit--test--repair session, followed by population management, defines the transition
\begin{equation}
    T_i:
    \big(\mathcal{P}_i,\mathcal{F}_i,\Omega_i\big)
    \longmapsto
    \mathcal{P}_{i+1},
    \label{eq:transition-operator}
\end{equation}
where
\begin{equation}
    \mathcal{F}_i = \{F_i^{(1)},\ldots,F_i^{(N_i)}\}
\end{equation}
is the record collection and \(\Omega_i\) fixes prompts, mutation rules, tool access and budgets. Thus proposals depend on observed behaviour, not only source text. Selection uses a declared subset
\begin{equation}
    D_i \subseteq \{1,\ldots,m_i\}
\end{equation}
through
\begin{equation}
    s_i(P)
    =
    \sigma_i\Big(
        \{L_i^\alpha(P)\}_{\alpha\in D_i},
        \mathcal{E}_i(P)
    \Big).
    \label{eq:stage-selection-score}
\end{equation}
where \(\sigma_i\) first applies the hard gates and then combines the decisive metrics. In the reported runs the evaluator and score are stationary, so the subscript records search step rather than a changing curriculum.

The transition \(T_i\) need not have one universal implementation: it may be a single completion, a targeted diff, a full rewrite, a genetic operator, or an agentic trajectory that edits, executes, diagnoses and repairs. These choices change how proposals are produced, but not the evaluator's role in defining admissibility and selection pressure. In the present runs, \(T_i\) is a coding-agent edit--test--repair session whose permitted mutation form is recorded in Table~\ref{tab:search-protocols}.

\paragraph{Island models.}
\label{subsec:island-models}

To preserve diversity, the archive is partitioned into islands,
\begin{equation}
    \mathcal{P}_i
    =
    \mathcal{I}_{i,1}\sqcup\cdots\sqcup\mathcal{I}_{i,K_i} .
    \label{eq:island-partition}
\end{equation}
where each \(\mathcal I_{i,k}\) maintains a partially independent lineage. All runs started with the same seed copied to two islands. The archived migration rate is zero: no candidate moved between islands. The structural and tree runs could instead spawn fresh islands after stagnation, giving four and three observed islands respectively. More generally, islands can preserve independent lineages, exchange candidates through migration, or restart exploration after stagnation; in the present runs diversity instead comes from independent archives, novelty filtering, prompt variation, and dynamic island creation.

\paragraph{Concrete protocols.}

Table~\ref{tab:search-protocols} records the executable protocol. A ``generation'' is one submitted mutation followed by one authoritative evaluation; the seed copies at generation zero are not counted as generated candidates. The bounded per-island archive is the active pool from which parents and prompt examples are drawn, while the database retains the full lineage. An ``inspiration'' is an additional archived program, with its feedback, shown to the coding agent alongside the selected parent: archive inspirations are sampled broadly from surviving programs, whereas top inspirations are drawn from the highest-scoring programs. They guide the proposal but are not combined mechanically. Items absent from the archive are marked NR.

\begin{table}[p]
\centering
\scriptsize
\setlength{\tabcolsep}{3pt}
\renewcommand{\arraystretch}{1.15}
\begin{tabularx}{\textwidth}{>{\raggedright\arraybackslash}p{0.16\textwidth}XXX}
\toprule
Protocol item & Scaling & Structural cancellation & Tree complexity \\
\midrule
Model; reasoning & \texttt{codex@gpt-5.6-sol}; medium (exact launcher identifier archived) & same mutation model; medium; separate local \texttt{gpt-5.5} strategist every two generations & same mutation model; medium reasoning effort \\
Interface and seed & \texttt{compute\_amplitude(plist)}; BCFW Python seed & \texttt{compute\_form(plist)}; CSW/MHV-vertex term seed & \texttt{amplitude\_M2(...)}; transparent multi-family BG/Dirac engine \\
Population/archive & 2 islands; 20 retained programs/island; prompts add 2 archive-sampled + 1 top-scoring inspiration & 2 initial islands, 4 observed after stagnation; 20/island; prompts add 2 + 1 inspirations & 2 initial islands, 3 observed; 30/island; prompts add 3 + 2 inspirations \\
Parent and acceptance & Within-island weighted sampling: sigmoid of robustly normalised score (scale 10), penalised by prior child count. Correct candidates enter a bounded archive; replacement requires beating its worst entry. & Same & Same \\
Migration/diversity & Migration rate 0; embedding novelty threshold 0.95, at most 2 novelty attempts & Migration rate 0; threshold 0.95; dynamic island after 5 stagnant generations & Migration rate 0; no embedding filter; dynamic island after 8 stagnant generations \\
Mutation and prompt & Full rewrite only; task prompt exposes interface, gates, timing grids, score and practice commands & Full rewrite only; interface, structural gates and conditioning score; custom diversity/meta prompts & 70\% diff, 30\% full rewrite; interface, counted operations, suite and self-test command \\
Visible/readable & Candidate, NumPy helpers, scoring formula and fresh-kinematics practice ruler; frozen exam points/reference table hidden & Candidate, practice ruler and physical-pole table; private solution engines and amplitude-zero exam pool hidden & Candidate and complete self-test evaluator in container; fresh correctness seeds generated at invocation \\
Writable/tools & Candidate and scratch work directory; shell and Python execution in Docker; no file/network access from submitted program & Same & Evolution block and scratch directory; shell/Python self-test; submitted code restricted to NumPy/math/itertools/cmath \\
Returned feedback & Gate failures; cell timings, fitted exponents and score components & Gate failures; per-configuration conditioning, terms and zero-pool diagnostic & Gate failures; per-process counts and ratios; runtime retained privately \\
Limits/stopping & 1 h/job; 8192 output tokens; 2 resamples, 3 patch attempts; 50 configured mutations; \$50 cap & 1 h/job; 8192; same resample/attempt caps; 20 configured, 18 submitted; \$50 cap & 3 h/job; 32768; same caps; manually resumed 20 to 50 configured; nominal \$2000 cap \\
Archive facts & 49 submitted; \$48.81 equivalent; 4.19 agent h & 18 submitted; \$14.54; 1.49 agent h & 48 submitted; \$49.14; 2.47 agent h \\
Randomness/human action & Frozen cell seeds; fresh gate seeds; master search seed NR. Infrastructure fixed before this released rerun; no mid-run edit recorded. & Fixed score/gate seeds and hidden zero pool; master seed NR. Pre-run meta/runtime patches; no mid-run edit recoverable. & Count point seed 12345; fresh secret gate seeds; master seed NR. Manual resume and automatic island spawn; no within-candidate human edit. \\
\bottomrule
\end{tabularx}
\caption{Concrete protocols for the released searches. Exact prompt text, visible-file manifests and candidate records are stored in the versioned repository~\cite{GuKrippendorf:2026repository}. ``NR'' means not recoverable from the archive. API-equivalent costs are accounting estimates under subscription access.}
\label{tab:search-protocols}
\end{table}

The master controller seed is not recoverable from the archived metadata. All submitted candidates, prompts, evaluator outputs and lineage relationships are nevertheless frozen in the released database, so the reported trajectories are exactly inspectable but cannot be regenerated stochastically from the initial controller state alone.

\paragraph{Cost of search.}
The released databases contain 115 submitted candidates, \(\$112.49\) of API-equivalent usage and 8.15 hours of active agent-session time. These are accounting estimates, not marginal spend under the subscription used. Of the 115 submissions, 114 passed every correctness gate. This high submission-level pass rate reflects the agentic proposal mechanism: the coding agent executes visible local tests and repairs failed edits before submitting a candidate to the authoritative evaluator. It does not mean that the internal edit--test trajectories were failure free, nor that the frozen correctness gates were unnecessary. The figures exclude exploratory and later continuation runs.

Beyond the three released trajectories, we performed exploratory searches with different objectives, prompts, population settings and computational budgets. Across these heterogeneous runs we repeatedly observed transitions toward reusable-current or helicity-adapted representations, R-invariant-style organisation, and shared recursive or symmetry-reduced matrix-element programs. Because the evaluators, scores and budgets differed, we treat these observations as qualitative evidence of recurrence rather than as statistically exchangeable repetitions.

\paragraph{Benchmark families used below.}
Two benchmarks concern tree-level colour-ordered gluon amplitudes: one evolves runtime scaling from a BCFW seed, while the other evolves the NMHV term decomposition under a cancellation-conditioning objective. The third evolves a transparent tree-level matrix-element engine over twenty QCD and electroweak processes using exact arithmetic count. Loop-level and end-to-end generator extensions remain future work.

\section{Three Probes of the Amplitude-Program Space}
\label{sec:results}

The experiments below are not three unrelated optimisation exercises. Together they probe whether self-evolving program search can traverse qualitatively different parts of the amplitude-program space: changes of theoretical representation, changes in the analytic organisation of intermediate singularities, and changes in the arithmetic structure of generator-relevant matrix-element programs.

Their common result is that external selection can induce conceptual transitions rather than only local implementation improvements. The successful lineages move between mathematical representations and compose structures that are normally developed separately. The objectives act as experimental dials: a scaling objective favours transferable and reusable structure, a cancellation objective favours more local analytic organisation, and an exact-operation objective favours combinations of recursion, symmetry, basis reduction, and shared computation. Table~\ref{tab:results-overview} gives the corresponding objectives; all candidates first pass numerical-correctness gates.

\begin{table}[t]
\centering
\small
\setlength{\tabcolsep}{3pt}
\begin{tabular}{p{0.18\linewidth}p{0.22\linewidth}p{0.27\linewidth}p{0.23\linewidth}}
\hline
Experiment & Evolved object & Decisive objective & Headline result \\
\hline
BCFW scaling & colour-ordered tree-gluon evaluator & \(0.75S_{\rm scaling}+0.25S_{\rm speed}\) & fixed-\(k\) split-helicity transfer path with \(O(n)\) source work; \(805\times\) geometric-mean speed-up \\
Cancellation structure & explicit NMHV rational form & \(S_{\rm pole}\) of Equation~\eqref{eq:structural-score} & score \(0.3758\to0.6048\); \(8\)--\(13\to2\)--\(3\) terms \\
Tree complexity & incoming-state-averaged \(\overline{|\mathcal M_n|^2}\) engine & geometric mean of baseline/candidate exact real-operation counts & \(195\to4\) million operations; score \(11.68\) \\
\hline
\end{tabular}
\caption{Overview of the three searches. Here \(Q_{0.95}\) is the empirical 95th percentile over a fixed phase-space sample. The tree-complexity score \(11.68\) is the geometric mean of per-process improvements, whereas \(47.7\times\) is the reduction of the summed operation count. The same champion gives a post-hoc \(5.9\times\) Python suite-runtime improvement.}
\label{tab:results-overview}
\end{table}

\subsection{From BCFW recursion to a linear-cost split-helicity evaluator}
\label{subsec:results-gluon}

\textbf{A scaling objective induces a change of mathematical representation.}

The first run evaluates tree-level colour-ordered gluon amplitudes \(A_n(1^{h_1},\ldots,n^{h_n})\), starting from a direct BCFW recursion. A timing cell fixes two integers: the number \(k\) of negative-helicity gluons and the total multiplicity \(n\). The score uses 34 cells: \(k=3\) at \(n=20,24,\ldots,64\); \(k=4\) at \(n=16,20,\ldots,40\); \(k=5\) at \(n=16,18,\ldots,30\); and \(k=6\) at \(n=14,16,\ldots,26\). A separate 19-case gate includes MHV, extreme and scattered-helicity configurations.

Numerical values are compared with a separately written Berends--Giele implementation, itself checked against Parke--Taylor amplitudes in the MHV sectors. A separate 19-case gate uses phase-space points and helicity arrangements disjoint from the 34 timing cells, including non-contiguous negative-helicity patterns, so correctness is not inferred from the configurations used for performance selection. The base tolerance is \(10^{-6}\). The archived protocol conservatively loosens the relative tolerance by multiplicity: \(10^{-5}\) above \(n=24\), \(10^{-4}\) above \(n=36\), \(10^{-3}\) above \(n=48\), and \(3\times10^{-3}\) above \(n=56\). An independent 40-digit \texttt{mpmath} recomputation of all 34 stored truths bounds the float64 reference error by \(1.0\times10^{-8}\) in the worst cell, at \(n=64\), so every rung has at least three orders of magnitude of headroom. The champion's worst error against this high-precision anchor is \(9.8\times10^{-12}\). The ladder is therefore a conservative guard and does not affect the reported result. A wrong value at any scored cell gives zero score.

For sector \(k\), a least-squares fit \(T(n)\sim n^{p_k}\) defines
\begin{equation}
 S_{\rm scaling}=\frac14\sum_{k=3}^{6}
 \log\!\left(1+\frac{p_k^{\rm seed}}{\max(p_k,0.25)}\right),
 \qquad
 S_{\rm speed}=\log(1+g),
\end{equation}
where \(g\) is the geometric mean of the 34 ratios \(T_{\rm seed}/T_{\rm candidate}\), and all logarithms in this score are natural. The frozen seed exponents are \(2.7975,3.6976,4.8928,6.4105\). The floor prevents near-flat, overhead-dominated fits from receiving an unbounded reward and gives \(S_{\rm scaling}\leq2.8916\); after that ceiling is reached, the speed term continues to resolve constant-factor gains. The selected score is \(0.75S_{\rm scaling}+0.25S_{\rm speed}\), encoding a search primarily for scaling while retaining speed as a tie-breaker.

\begin{figure}[t]
  \centering
  \includegraphics[width=\linewidth]{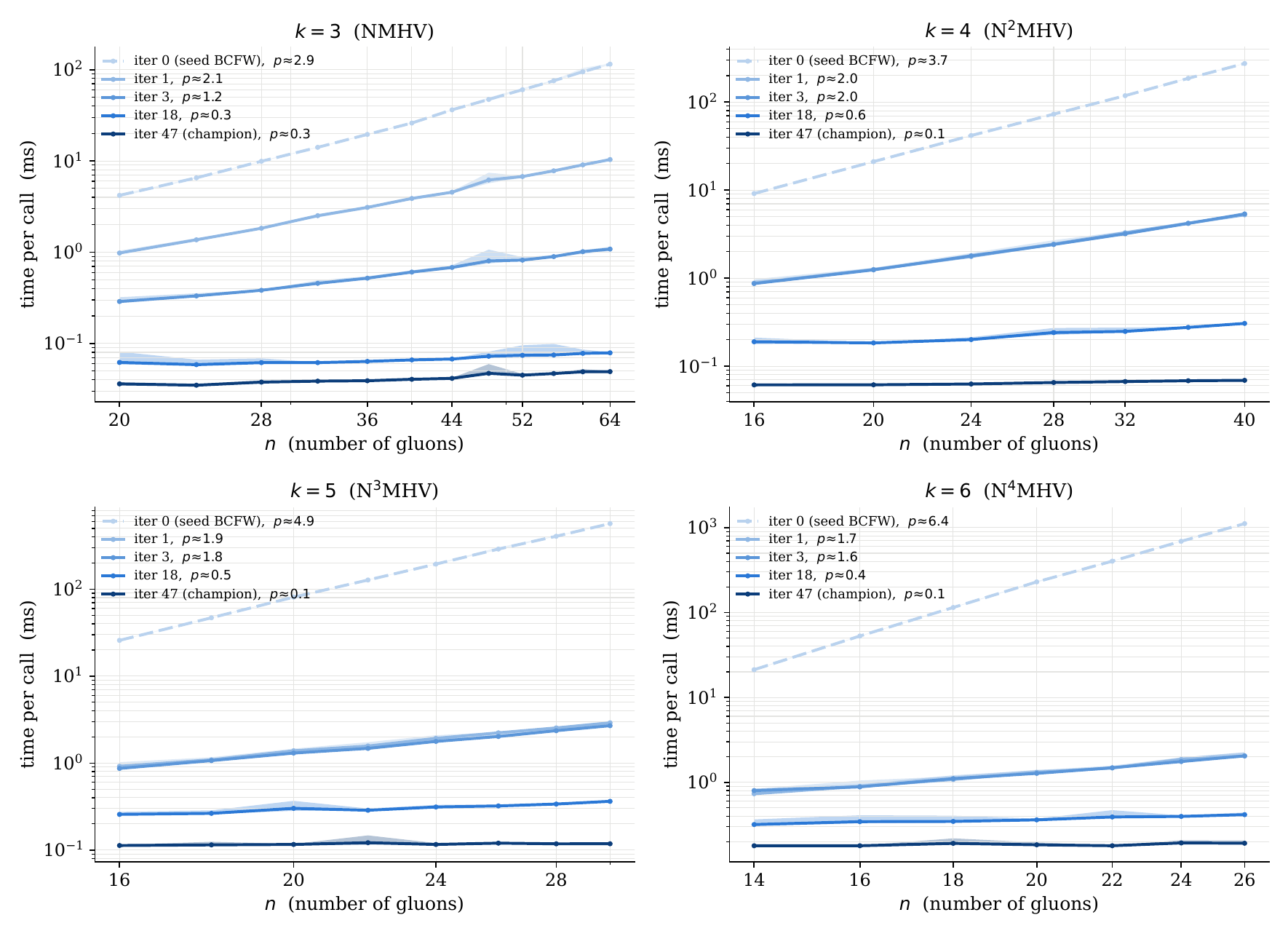}
  \caption{Multiplicity scaling along the completed fifty-generation run. Each point is the median wall-clock time per call over three phase-space points and five repeats per point; shaded bands show the interquartile range and are often narrower than the line width. These post-hoc measurements used Python~3.11 and NumPy~2.4.6. Fitted slopes are insensitive to using minima, medians or means, differing by at most a few times \(10^{-2}\). The generation-18 representation change introduces the fixed-\(k\), linear-cost split-helicity path; the generation-47 champion further improves constant factors. Near-flat curves on the scored windows are dominated by fixed NumPy overhead.}
  \label{fig:gluon-iteration-improvement}
\end{figure}

In the released fifty-generation run, the generation-47 champion scores \(3.8416\), compared with \(0.6817\) for the seed, and reaches \(g=804.7\), as summarised in Figure~\ref{fig:gluon-iteration-improvement}. It saturates the scaling ceiling: \(3.8416=0.75(2.8916)+0.25\log(1+804.7)\) to the reported precision. Its selection-time fitted exponents over the scored windows are \(0.249,0.048,0.005,0.107\), down from \(2.95,3.71,4.90,6.47\) in that run. These near-zero fitted values are overhead-dominated measurements rather than a claim of sublinear asymptotic complexity: a sweep to \(n=256\) exposes the underlying linear growth.

The lineage first moves from BCFW to reusable Berends--Giele currents and a CSW/MHV-vertex organisation, then reaches a direct implementation of the known split-helicity zigzag sum~\cite{Britto:2005ha}. The generation-47 champion reorganises that sum as a finite-state transfer dynamic program. Its four-component numerator state uses the cubic basis \((x^3,x^2y,xy^2,y^3)\), i.e. the symmetric-cube representation of the two-component spinor space under the general \(2\times2\) interval maps appearing in the formula. Reverse suffix scans factor repeated endpoint sums. The source-level work is \(O(k^2n)\), hence \(O(n)\) at fixed negative-helicity count \(k\), with \(O(n)\) memory.

For the scored contiguous patterns, static call-graph inspection confirms that \texttt{compute\_amplitude} returns this specialised path directly; no fallback contributes. Scattered patterns retain CSW or Berends--Giele routes and are not covered by the linear fixed-\(k\) claim. Fresh tests at \(n=96,128\) and timing through \(n=256\) support the extrapolation beyond the fitted window, while Appendix~\ref{app:robustness} tests route and timing stability across unseen kinematics. The frozen candidate and evaluator are in the \href{https://github.com/YiGu310/scattering-amplitudes-program-search/tree/ab8c8682713ff8592bb55e456229619bff7cfd12/experiments/1_bcfw_scaling}{versioned scaling archive}~\cite{GuKrippendorf:2026repository}.

\subsection{Evolving representations with less spurious cancellation}
\label{subsec:results-poles}

\textbf{A structural objective reorganises where unphysical cancellation occurs.}

The second run asks a different physics question: can search find an equivalent expression in which unphysical singularities cancel more locally? A candidate program \(P\) returns an explicit rational form \(A_n(P;\Phi)=\sum_{t=1}^{N_{\rm term}}T_t(P;\Phi)\), where \(T_t=N_t/\prod_{f=1}^{N_t^{(D)}}D_{tf}\), for scattered-helicity NMHV gluon amplitudes. Here \(t\) labels terms and \(f\) labels denominator factors within a term. The four scored configurations are \(n=6\) with negative legs \((0,2,4)\) and \((0,1,3)\), and \(n=7\) with negative legs \((0,2,4)\) and \((0,1,4)\); an \(n=8\) configuration is used only as a correctness gate.

Individual terms may contain spurious poles that cancel only in the complete amplitude. Near such a boundary, large terms can sum to a modest finite answer. For a candidate representation \(P\) and phase-space point \(\Phi\), the evaluator uses
\begin{equation}
\begin{aligned}
 \kappa_{\rm sum}(P;\Phi)
 &=\frac{\sum_{t=1}^{N_{\rm term}}|T_t(P;\Phi)|}
 {|\sum_{t=1}^{N_{\rm term}}T_t(P;\Phi)|},
 &r_c(P)&=Q_{0.95}\!\left[\log_{10}\kappa_{\rm sum}(P;\Phi)\right],
 \\
 S_{\rm pole}(P)
 &=\left[1+\frac14\sum_{c=1}^{4}r_c(P)\right]^{-1}.&&
\end{aligned}
 \label{eq:structural-score}
\end{equation}
For a fixed term decomposition, \(\kappa_{\rm sum}\geq1\) is the condition number of summation under independent perturbations of the terms. Its logarithm is a proxy for sensitivity to inter-term cancellation, not the exact number of floating-point digits lost, which also depends on term evaluation and summation order. The metric is representation dependent by construction: it does not count poles and is not an amplitude invariant. For each configuration, the code takes NumPy's default linearly interpolated 95th percentile over 300 fixed points after dropping values whose amplitude magnitude is below \(10^{-3}\) of the finite-sample median. Six gates enforce numerical correctness, fixed term structure, finite declared numerators, a denominator-factor cap, irreducibility against weighted mixtures, and nontrivial term cancellation at independently certified amplitude zeros.

The CSW-style seed scores \(0.3758\); its diagrams expose reference-spinor-dependent denominators even though their sum is reference independent. Generation 1 removes that auxiliary spinor and reaches an NMHV R-invariant-style cell expansion, scoring \(0.5173\). Later candidates identify neighbouring cells with a common denominator factor and combine them pointwise into larger corridor terms. This executable gluing is consistent with the expected algebraic cancellation and is verified numerically along controlled boundary trajectories. The released implementation removes the shared denominator key after pointwise division, but does not provide a symbolic polynomial-divisibility certificate. The champion scores \(0.6048\), replacing 8--13 seed terms by 2 terms at six points and 3 at seven points, with \(r_c\) reduced from \(1.47\)--\(2.12\) to \(0.40\)--\(0.91\). It remains short of the spurious-free ideal because outer corridor boundaries remain.

We additionally tested the interpretation along controlled momentum-twistor trajectories approaching named spurious boundaries. At the shared six-point boundary removed by the champion, the two unglued cells develop equal-and-opposite \(1/S\) residues and \(\kappa_{\rm sum}\propto |S|^{-1}\), whereas their glued supercell remains \(O(1)\) and its summation error stays at working-precision round-off. At a boundary retained by the champion the \(|S|^{-1}\) sensitivity remains. This boundary-specific control shows that the score improvement reflects removal of the identified internal boundary rather than a universal rescaling of the terms.

Hodges' momentum-twistor construction is the analytic endpoint of the comparison: for NMHV amplitudes it makes possible representations without spurious poles and with manifest geometric organisation~\cite{Hodges:2009spurious,ArkaniHamed:2010polytope}. The champion is not that endpoint; it is a partial algebraic analogue based on component cells and local boundary gluing. The frozen run is in the \href{https://github.com/YiGu310/scattering-amplitudes-program-search/tree/ab8c8682713ff8592bb55e456229619bff7cfd12/experiments/2_spurious_structure}{versioned structural archive}~\cite{GuKrippendorf:2026repository}.

\begin{figure}[t]
  \centering
  \includegraphics[width=\linewidth]{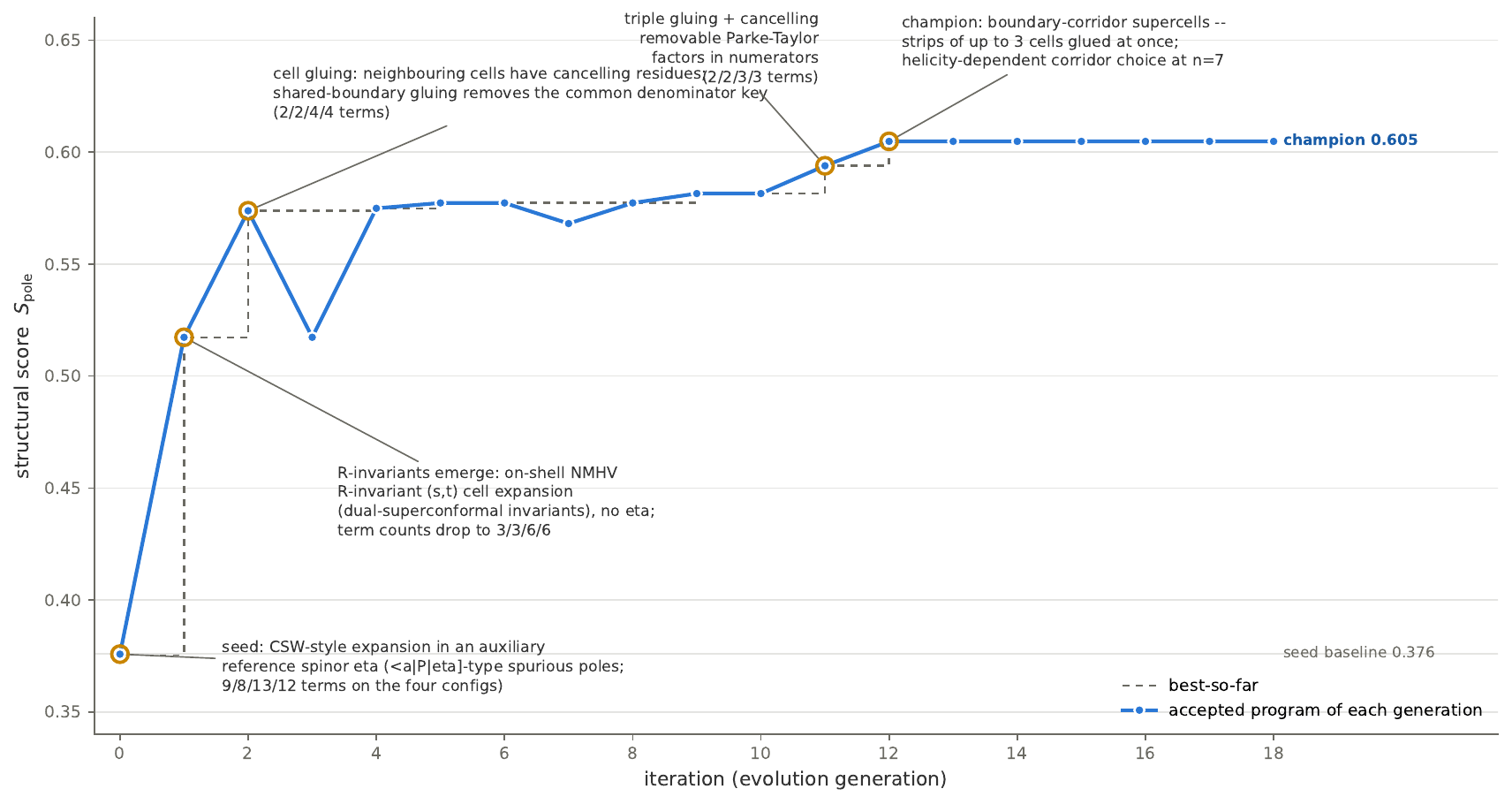}
  \caption{Evolution of the cancellation-conditioning score \(S_{\rm pole}\). Blue points show accepted candidates and the dashed staircase the best-so-far score; amber circles mark representation changes. The large first step changes from the CSW-style seed to an NMHV cell expansion. Subsequent gains combine shared-boundary cancellations into glued terms, ending at \(0.605\) compared with the seed's \(0.376\). Reproducibility metadata, including the internal run identifier, appear in Table~\ref{tab:search-protocols}.}
  \label{fig:pole-score-evolution}
\end{figure}

\subsection{Tree-amplitude complexity and production-tool comparison}
\label{subsec:results-generators}

\textbf{A phenomenological complexity objective composes recursion, symmetry, basis reduction, and shared computation into a hybrid matrix-element engine.}

This experiment asks how much arithmetic can be removed from a small but varied tree-level matrix-element engine. For each process the program forms the exact colour--helicity sum and applies the incoming spin--colour average of Equation~\eqref{eq:squared-matrix-element}. For identical final gluons, and for \(W^+W^-\to ZZ\), the archived scalar also divides by the corresponding final-state factorial; this is a separate phase-space symmetry convention, not part of the initial-state average. Individual engines were checked either directly against generated MadGraph references or against frozen project references previously validated point by point. The evolved candidate is an independent Python implementation: the search edits neither MadGraph nor Sherpa and does not cover phase-space integration, subtraction, matching, showering or unweighting.

The suite contains three scaling families---pure gluons at \(n=4,5,6\), one massless quark line at \(n=4,5,6\), and \(u\bar u\to Z+kg\) for \(k=1,\ldots,4\)---plus ten fixed processes spanning massive QCD, QED, charged and neutral currents, Yukawa interactions, two fermion lines, diboson and associated \(ZH\) production, and a quartic gauge vertex. The counted region begins after construction of the external spinors and polarisation vectors and ends when the averaged scalar \(\overline{|\mathcal M|^2}\) has been formed. This isolates the reusable matrix-element kernel rather than the entire event-generation workflow. The full process definitions and validation references are listed in Table~\ref{tab:processes} of Appendix~\ref{app:processes}.

The decisive metric is
\begin{equation}
 C_{\rm tree}(j)=N_{+}+N_{-}+N_{\times}+N_{/}+N_{\rm neg},
 \qquad
 S_{\rm tree}=\left[\prod_{j=1}^{20}
 \frac{C_{\rm seed}(j)}{C_{\rm candidate}(j)}\right]^{1/20},
\end{equation}
The first expression counts additions, subtractions, multiplications, divisions and unary negations on real numbers; complex arithmetic is expanded into this fixed convention. Here \(j\) labels one of the twenty processes; the ratio \(C_{\rm seed}(j)/C_{\rm candidate}(j)\) is its improvement factor. Their geometric mean \(S_{\rm tree}\) gives every process equal logarithmic weight, preventing the largest pure-gluon cases from determining the score by themselves. A strict counted-array type rejects unregistered operations, conversion to raw arrays, numerical branching and hidden powers. Before a score is assigned, each candidate is compared with the oracle on four fresh phase-space points for every process at relative tolerance \(10^{-6}\), and the value produced by the counted path is checked again. Wall-clock timings are recorded privately but are absent from the score and feedback.

\paragraph{Kinematic independence of the phenomenological objective.}
The search selects one program for the full process suite; it does not choose a method separately at each phase-space point. More strongly, the counted-array interface forbids branching on numerical kinematics, so each process has a fixed arithmetic graph and its exact operation count is phase-space independent by construction. We verified this explicitly for both seed and champion at 20 unseen points for each of the twenty processes: all 400 evaluations per engine reproduce the archived operation count bit for bit, while the worst relative error over the four process families is \(9.2\times10^{-14}\). Thus the \(47.7\times\) aggregate reduction is not an optimisation to one favourable kinematic point. Appendix~\ref{app:robustness} gives this audit in full and reports analogous fresh-point tests for the scaling and structural searches.

\subsubsection{Fifty-generation result}

\begin{figure}[t]
  \centering
  \includegraphics[width=\linewidth]{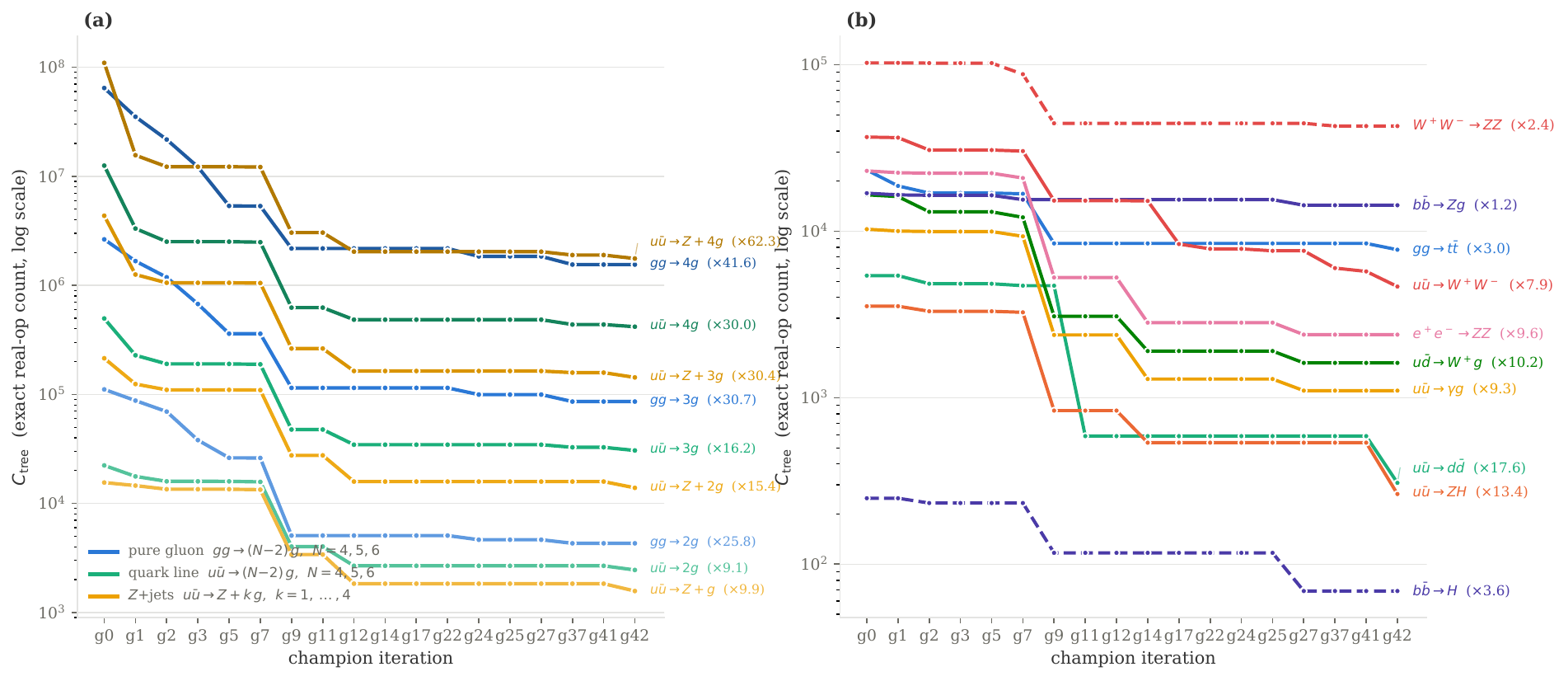}
  \caption{Exactly counted real arithmetic operations \(C_{\rm tree}\) for the twenty benchmark processes along the champion lineage. (a) Pure-gluon, single-quark-line and \(Z+\)jets scaling families; shades distinguish multiplicities. (b) Ten fixed-multiplicity processes covering the remaining interaction types. Line-end labels give the seed-to-champion improvement. Definitions and validation references appear in Table~\ref{tab:processes}; the internal run identifier is recorded in Table~\ref{tab:search-protocols}.}
  \label{fig:generators-iteration-improvement}
\end{figure}

Figure~\ref{fig:generators-iteration-improvement} summarizes the operation-count trajectory across the champion lineage.

The released run was configured to a 50-generation target and is the version published in the companion repository. Accepted candidate directories are labelled 0--48; the generation-42 champion remains unbeaten through the end of the archive. It improves \(S_{\rm tree}\) from 1 to \(11.680\). The unweighted suite total falls from \(195{,}010{,}588\) to \(4{,}087{,}932\) real operations, a factor \(47.7\). These two factors answer different questions: \(11.68\times\) is the selected geometric mean that gives every process equal logarithmic weight, whereas \(47.7\times\) is dominated by the largest processes. The largest individual gain is \(62.3\times\) for \(u\bar u\to Z+4g\); gains are \(41.6\times\) for six gluons and \(30.0\times\) for a six-particle massless quark-line process. The weakest gain, \(1.18\times\) for \(b\bar b\to Zg\), is consistent with the champion's reliance on massless chirality.

The conclusion is robust to the equal-weight convention in \(C_{\rm tree}\). Re-instrumenting the frozen seed and champion by operation type shows that divisions fall from \(2{,}384{,}628\) to 837 across the suite. Six named cost models and 1000 log-uniform random weight vectors with \(w_\times\in[1,4]\), \(w_{\div}\in[1,32]\), and \(w_{\rm neg}\in[0,1]\) give aggregate reductions of \(47.7\)--\(57.2\times\) and geometric-mean reductions of \(11.7\)--\(12.4\times\); none makes any process worse. The reported equal-weight count is therefore conservative within this family of arithmetic cost models. Appendix~\ref{app:processes} gives the reweighting study; it remains an operation-count analysis, not a replacement for wall-clock measurement.

Table~\ref{tab:tree-lineage} gives a code-level anatomy of the champion lineage. The seed already memoises Berends--Giele currents within one colour ordering; the first searched step shares them across orderings. Later candidates combine basis, helicity, parity, chirality and contraction changes. The operation totals are observations at lineage checkpoints, not ablations, so they establish temporal association rather than the isolated causal gain of any principle. Post-hoc wall-clock time for the whole suite falls from \(0.1399\) to \(0.0236\) seconds, a \(5.9\times\) improvement. Timing was neither selected nor shown to the agent.

\begin{table}[t]
\centering
\scriptsize
\setlength{\tabcolsep}{3pt}
\begin{tabularx}{\textwidth}{p{0.10\textwidth}p{0.21\textwidth}p{0.25\textwidth}p{0.18\textwidth}X}
\toprule
Generation & Added design principle & Code-level consequence & Main affected processes & Observed operation-count association \\
\midrule
Seed & memoised Berends--Giele currents & interval currents cached within each ordering & gluon, quark, \(Z+\)jets & baseline \(195.01\)M suite total \\
1--3 & sharing across colour orderings; reflection/parity reuse & common current cache and folded ordering orbits & scaling families & \(57.76\)M at g1; \(29.29\)M at g3 \\
5 & Kleiss--Kuijf basis reduction & fewer independent pure-gluon orderings & pure gluons & \(22.11\)M \\
7--11 & helicity-sector pruning and orbit reuse & vanishing sectors skipped; helicity-equivalent work shared & gluon and massless lines & \(21.95\)M at g7; \(6.41\)M at g11 \\
12--17 & two-component Weyl recursion; chirality/parity caching & four-component Dirac work replaced on massless paths & quark, \(Z+\)jets, fermion-line processes & \(5.14\)M at g12; \(5.12\)M at g17 \\
24--26 & direct top-level contractions & final currents contracted without materialising full objects & several fixed and scaling families & \(4.77\)M \\
27--41 & coupling factorisation and shared vector-boson currents & common scalar factors and emissions moved outside repeated contractions & fixed electroweak and fermion-line processes & \(4.77\)M at g27; \(4.26\)M at g41 \\
42 & Ward/Fierz-normalised contraction cleanup & algebraically reduced terminal kernels & mixed suite & champion \(4.09\)M \\
\bottomrule
\end{tabularx}
\caption{Conceptual transitions in the tree-complexity champion lineage. Checkpoint totals are associated with bundled candidate changes and must not be read as single-feature causal ablations.}
\label{tab:tree-lineage}
\end{table}

\subsubsection{Matched comparison with MadGraph and Sherpa-Comix}

The purpose of the MadGraph and Sherpa comparison is not to claim a universal ranking between the evolved Python engine and mature production generators. It is to calibrate whether the representation and arithmetic changes found by search are large enough to matter in a generator-relevant operating regime. At \(n=6\), the searched implementation is \(277\times\) faster than the tested Sherpa-Comix exact-colour, exact-helicity call and \(13.7\times\) slower than process-specialised MadGraph Fortran compiled at \texttt{-O3}. Together with the \(47.7\times\) aggregate reduction in counted arithmetic, this indicates that large speed-ups remain available when search can combine colour organisation, helicity structure, recursive reuse, basis reduction, and process specialisation.

For pure-gluon scattering at \(n=4,5,6\), the seed, champion, MadGraph and Sherpa-Comix were evaluated on the same pre-generated phase-space points and the same full-colour, helicity-summed and incoming-state-averaged \(\overline{|\mathcal M_n|^2}\), including the same final-state symmetry normalisation. One-time initialisation is excluded for every implementation. The timed calls are the momentum-to-\(\overline{|\mathcal M_n|^2}\) call for the Python programs, \texttt{py\_get\_value} inside the MadGraph worker, and \texttt{CSMatrixElement()} in a dedicated Sherpa-Comix driver. At the first point, all four outputs agree to relative precision between \(10^{-13}\) and \(10^{-9}\), depending on multiplicity and implementation.

The MadGraph benchmark uses MG5\_aMC v3.6.7 with \texttt{output standalone} and a process-specific \texttt{matrix2py} interface to compiled Fortran. To separate the generated algorithm from compiler acceleration, one frozen generated-source snapshot was rebuilt at \texttt{-O0}, \texttt{-O2}, and \texttt{-O3}; only the gfortran optimisation flag changed, and the HELAS, model, matrix-element, linked-library, and f2py layers were rebuilt consistently. All optimisation levels agree in \(\overline{|\mathcal M|^2}\) to at worst \(1.2\times10^{-11}\). The Sherpa benchmark uses Sherpa v3.0.4, built in CMake Release mode with AppleClang 15.0.0.15000309 and \texttt{-O3 -DNDEBUG}; optional LHAPDF, HepMC3, FastJet, Rivet, Pythia and OpenLoops support was disabled. The Python stack used Python 3.11.15 and NumPy 1.26.4. The driver fixes \(\alpha_s=0.118\), zero widths and a diagonal CKM matrix, uses \(gg\to2g,3g,4g\), and excludes inter-process communication and one-time initialisation.

\begin{table*}[t]
\centering
\small
\renewcommand{\arraystretch}{1.18}
\setlength{\tabcolsep}{7pt}
\begin{tabular}{c r r r r r r}
\toprule
\(n\) & \multicolumn{1}{c}{\shortstack{Seed\\Python}} & \multicolumn{1}{c}{\shortstack{Champion\\Python}} & \multicolumn{1}{c}{\shortstack{MG\\\texttt{-O0}}} & \multicolumn{1}{c}{\shortstack{MG\\\texttt{-O2}}} & \multicolumn{1}{c}{\shortstack{MG\\\texttt{-O3}}} & \multicolumn{1}{c}{\shortstack{Sherpa\\\texttt{-O3}}} \\
\midrule
4 & \(548.2\,\mu{\rm s}\) & \(295.9\,\mu{\rm s}\) & \(8.38\,\mu{\rm s}\) & \(3.61\,\mu{\rm s}\) & \(3.81\,\mu{\rm s}\) & \(3.031\,{\rm ms}\) \\
5 & \(4.810\,{\rm ms}\) & \(1.141\,{\rm ms}\) & \(124.95\,\mu{\rm s}\) & \(47.94\,\mu{\rm s}\) & \(52.32\,\mu{\rm s}\) & \(102.493\,{\rm ms}\) \\
6 & \(110.710\,{\rm ms}\) & \(14.470\,{\rm ms}\) & \(3.076\,{\rm ms}\) & \(0.986\,{\rm ms}\) & \(1.056\,{\rm ms}\) & \(4.006\,{\rm s}\) \\
\bottomrule
\end{tabular}
\caption{Matched median per-point evaluation times for \(gg\to(n-2)g\), excluding one-time setup. The same generated MadGraph v3.6.7 Fortran source was compiled at each stated optimisation level; Sherpa v3.0.4 used a Release \texttt{-O3} build. At least twenty timing batches were used per point. These are implementation-level measurements on one software/hardware stack, not universal properties of MadGraph or Sherpa.}
\label{tab:mg-sherpa-timing}
\end{table*}

\begin{figure}[t]
  \centering
  \includegraphics[width=0.90\linewidth]{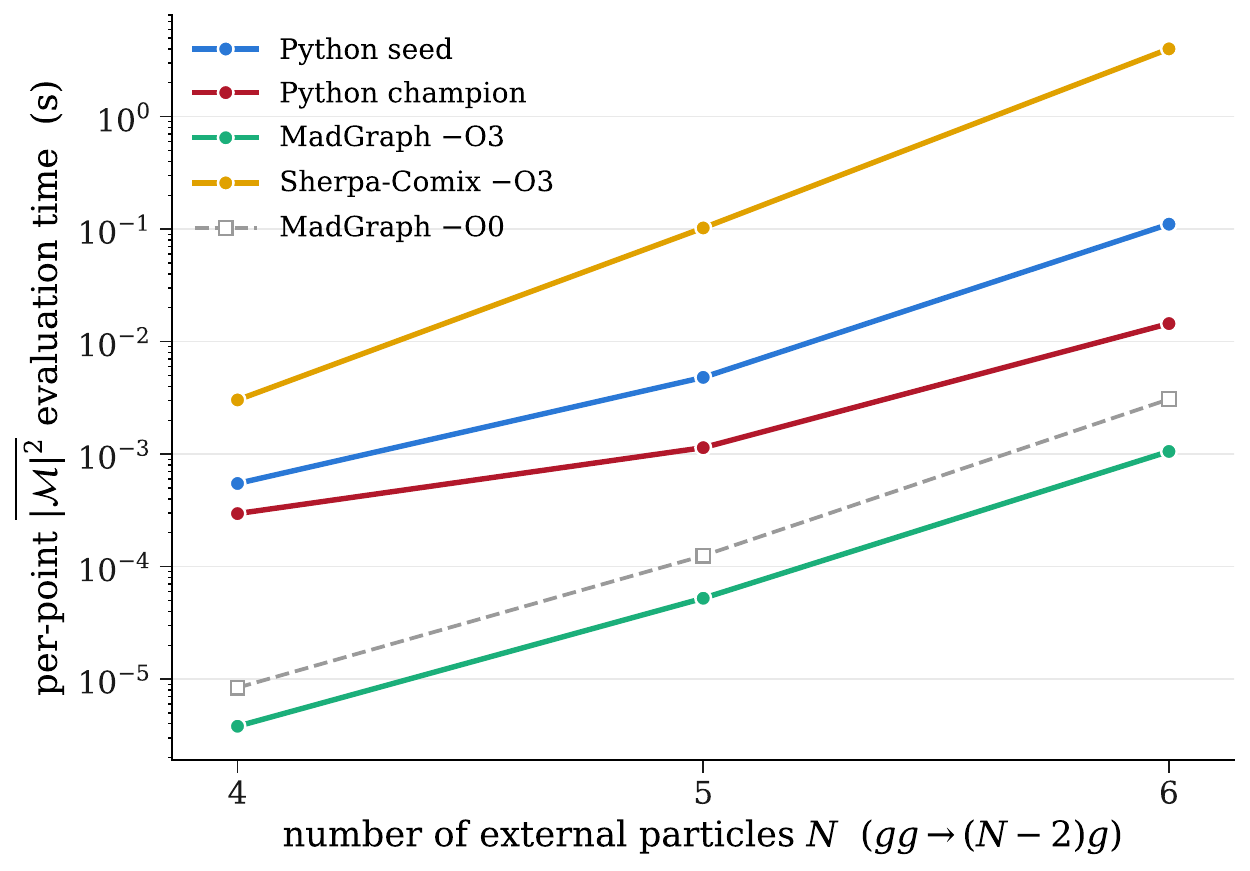}
  \caption{Compiler-controlled pure-gluon matrix-element timing comparison for \(gg\to(n-2)g\) at \(n=4,5,6\). The Python seed and generation-42 champion are compared with MadGraph \texttt{matrix2py} built at \texttt{-O3} and Sherpa-Comix built in Release mode at \texttt{-O3}; MadGraph \texttt{-O0} is retained as a faint reference. All curves use identical pre-generated points, the full-colour, helicity-summed and incoming-state-averaged \(\overline{|\mathcal M_n|^2}\), the same final-state symmetry normalisation, and exclude one-time setup. Rebuilding the same MadGraph source at \texttt{-O0}, \texttt{-O2}, and \texttt{-O3} changes the result by at most \(1.2\times10^{-11}\). Wall time was unavailable to the search.}
  \label{fig:mg-sherpa-timing}
\end{figure}

The champion is \(10\times\), \(90\times\), and \(277\times\) faster than Sherpa-Comix at \(n=4,5,6\), respectively. MadGraph remains faster in absolute time, but against \texttt{-O3} the gap narrows from \(77.7\times\) at \(n=4\) to \(13.7\times\) at \(n=6\). Recompiling the same MadGraph source from \texttt{-O0} to \texttt{-O3} gives \(2.2\)--\(2.9\times\) acceleration; \texttt{-O2} is marginally faster than \texttt{-O3} on this stack. These are the directly tested implementations displayed in Figure~\ref{fig:mg-sherpa-timing}. The finite-multiplicity trend is favourable to the evolved engine, but it should not be extrapolated beyond \(n=6\) without new measurements and does not include phase-space integration, event unweighting, matching, merging or showering.

\subsubsection{Interpretation relative to existing generators}

The discovered ingredients are not uniformly absent from established generators. Comix is itself based on colour-dressed Berends--Giele recursion~\cite{Gleisberg:2008fv}; cached off-shell currents are therefore part of its basic organisation. MadGraph instead generates process-specific helicity-amplitude code and already benefits from compiled specialisation and common wavefunction reuse~\cite{Alwall:2014hca}, as well as helicity recycling~\cite{Mattelaer:2021helicity,HSFGenerators:2021cpu}. Both tools also organise colour and helicity nontrivially. The result should not be described as introducing recursion, helicity reuse or colour structure to programs that lacked them.

The evolved engine combines these ideas in a different computational representation. Its pure-gluon path uses colour-ordered currents, a Kleiss--Kuijf-reduced set of orderings, reversal and parity relations, shared current caches, and an aggregated exact colour contraction. Comix instead carries colour through a colour-dressed recursion, while standard MadGraph generated code is organised around diagrams and HELAS/ALOHA wavefunctions. Consequently, absence of an explicit Kleiss--Kuijf basis in Comix is not simply a missed local optimization: it reflects a different colour representation.

The comparison also selects a particular operating regime. The benchmark requires an exact full-colour and full-helicity sum, whereas production event generation can use colour and helicity sampling, an important design point for Comix. The \(277\times\) figure is therefore a result for the isolated exact-sum kernel and tested interface, not a universal Sherpa slowdown or an event-throughput comparison. The important conclusion is not that the present Python engine should replace either generator. It is that the observed gains are sufficiently large to justify transferring the discovered transformations into native code-generation and event-generation workflows. Native integration is the next decisive experiment.

The frozen evaluator, lineage, exact-operation counter and timing harness are in the \href{https://github.com/YiGu310/scattering-amplitudes-program-search/tree/ab8c8682713ff8592bb55e456229619bff7cfd12/experiments/3_tree_complexity}{versioned tree-complexity archive}~\cite{GuKrippendorf:2026repository}.

\section{Discussion and Conclusions}
\label{sec:discussion}

The principal result is that repository-scale self-evolving agents can search conceptually over scattering-amplitude programs. The successful lineages change representation and assemble hybrid algorithms from recursion, analytic identities, symmetry, basis reduction, dynamic programming and shared computation. They do more than alter constants or local syntax.

The scaling and structural searches reproduce a familiar lesson from amplitude theory: an apparently difficult calculation can become simpler when its representation changes. The novelty is not the prior existence of split-helicity formulae, R-invariants, CSW or Berends--Giele recursion. Nor do we claim that these concepts were absent from model pretraining. The empirical result is that a fixed external objective selects and composes them into executable organisations without being supplied a target implementation. The archived programs and numerical audits make the resulting computational claims inspectable while separating known amplitude identities from searched code structure.

The phenomenological result is concrete. Exact-operation selection reduces the twenty-process aggregate by \(47.7\times\) and the hidden post-hoc Python runtime by \(5.9\times\); the reduction remains at least \(47\times\) under the tested operation-cost reweightings. At \(n=6\), the evolved exact-sum gluon kernel is \(277\times\) faster than the tested Sherpa-Comix call and approximately \(15\times\) slower than the fastest tested optimised MadGraph Fortran build. This is not a universal generator ranking: the implementations use different representations, Comix can sample colour and helicity, and production includes phase-space integration, subtraction, unweighting, matching, merging and showers. It shows that the discovered transformations are large enough to warrant native implementation and end-to-end tests.

The kinematic audit in Appendix~\ref{app:robustness} addresses adaptation to sampled phase space. The specialised scaling route is unchanged and remains the fastest tested lineage member in 340 fresh cell--point cases; operation counts are invariant across 400 fresh process--point evaluations for seed and champion; and the structural improvement persists on 1000-point holdouts per helicity configuration. We find no evidence of kinematic overfitting in these tests. The process-level statement is strongest because numerical branching is rejected by construction, making the \(47.7\times\) exact-operation reduction point independent.

Important limits remain. The audits do not establish transfer to arbitrary helicities, processes or multiplicities. The controlled spurious-boundary trajectory provides high-precision numerical evidence for the named residue cancellation but is not a symbolic divisibility proof. The process suite is small compared with the Standard Model process space, and the generator comparison isolates one kernel on one stack. An evolved program is evidence, not explanation; source-level interpretation, independent validation and native integration are required before it becomes a production optimisation.

All wall-clock measurements used a 14-inch 2021 MacBook Pro (MacBookPro18,3), Apple M1 Pro with six performance and two efficiency cores, 16~GB unified memory, and macOS 26.0.1 (build 25A362), on mains power. Background load was not logged; selection used best-of-repeats timing, while Figure~\ref{fig:gluon-iteration-improvement} reports medians and interquartile ranges. The \href{https://github.com/YiGu310/scattering-amplitudes-program-search/tree/ab8c8682713ff8592bb55e456229619bff7cfd12}{versioned companion repository} stores the frozen runs and reproduction material~\cite{GuKrippendorf:2026repository}.

The conclusion is not that any champion is final. It is that the physical answer can remain fixed while the executable route forms a large, structured and searchable design space. Scaling, cancellation and exact-operation objectives select different useful organisations, showing that substantial gains remain above the level of low-level tuning.

The next decisive step is practical: transfer the strongest transformations into native generator code and measure realistic phase-space and end-to-end event throughput. That programme requires amplitude theorists, phenomenologists, generator developers, numerical experts and program-search researchers. Experts define trustworthy objectives and standards of evidence; self-evolving systems expand the scale on which representations and combinations can be explored. We hope that the coding-agent-based program search introduced and demonstrated here will help shape amplitude understanding and enable efficient analyses of high-luminosity LHC data.

\section*{Use of AI systems}
OpenAI Codex was used as the proposal mechanism in the externally evaluated program-search loops described in Section~\ref{sec:search-design}. Agent-generated candidates were evaluated by frozen, author-designed correctness tests and objective functions before entering the program population. The model weights, evaluators, and objectives remained fixed during each reported run. The authors designed the scientific tasks and validation protocols, inspected and interpreted the resulting programs, and independently verified the claims reported here.

ChatGPT, Codex, Anthropic Claude, and Fable were additionally used for code assistance, manuscript editing, and figure preparation. All AI-generated material was reviewed and revised by the authors, who take full responsibility for the content and conclusions.

\section*{Acknowledgements}
S.K. thanks Ulrich Haisch, Tilman Plehn, Gary Shiu, and David Skinner for discussions, and Michael Spannowsky for discussions and comments on an early draft of this paper. S.K. has been partially supported by STFC consolidated grants ST/T000694/1 and ST/X000664/1.

\clearpage
\appendix
\section{Twenty-process tree-complexity suite}
\label{app:processes}

This appendix defines the process suite used by the exact-operation objective in Section~\ref{subsec:results-generators}. The suite combines multiplicity-scaling families with fixed-multiplicity processes chosen to expose distinct QCD and electroweak interaction structures; the corresponding frozen engines and validation data are available in the companion repository.

\subsection{Benchmark conventions}

The normalisation factor \(\mathcal N_n\) in Equation~\eqref{eq:single-trace-decomposition} is one when the partial amplitude is defined in the same trace convention. The frozen pure-gluon engine instead computes a bare Berends--Giele partial amplitude \(A_{\rm BG}\); its MadGraph-normalised ordered coefficient has magnitude \(2g_s^{n-2}A_{\rm BG}\), up to one common phase that cancels after squaring. This is the executable conversion used in the matched benchmark.

The tree-complexity engine forms \(\Sigma_n\) and divides by the incoming spin and colour factors to obtain \(\overline{|\mathcal M_n|^2}\). For processes with identical final particles its archived scalar additionally includes \(1/S_{\rm ident}\): the corresponding final-state factorial for identical final gluons and for \(W^+W^-\to ZZ\). This is a phase-space symmetry convention applied after the incoming-state average, not part of the overline definition.

\begin{table}[h]
  \centering
  \scriptsize
  \setlength{\tabcolsep}{2.5pt}
  \begin{tabular}{llllrl}
    \hline
    Label & Process & Interaction content & \(n\) & Baseline \(C_{\rm tree}\) & Validation reference \\
    \hline
    \multicolumn{6}{l}{\emph{Scaling families}}\\
    \texttt{gluon\_n4} & \(gg\to2g\) & pure-gluon QCD & 4 & 111{,}093 & project BG engine (MadGraph-endorsed) \\
    \texttt{gluon\_n5} & \(gg\to3g\) & pure-gluon QCD & 5 & 2{,}641{,}289 & project BG engine (MadGraph-endorsed) \\
    \texttt{gluon\_n6} & \(gg\to4g\) & pure-gluon QCD & 6 & 64{,}567{,}497 & project BG engine (MadGraph-endorsed) \\
    \texttt{quark\_n4} & \(u\bar u\to2g\) & single massless quark line & 4 & 22{,}302 & project engine (MadGraph-endorsed) \\
    \texttt{quark\_n5} & \(u\bar u\to3g\) & single massless quark line & 5 & 496{,}714 & project engine (MadGraph-endorsed) \\
    \texttt{quark\_n6} & \(u\bar u\to4g\) & single massless quark line & 6 & 12{,}534{,}584 & project engine (MadGraph-endorsed) \\
    \texttt{zjet\_k1} & \(u\bar u\to Z+g\) & neutral current + QCD & 4 & 15{,}536 & project engine (MadGraph-endorsed) \\
    \texttt{zjet\_k2} & \(u\bar u\to Z+2g\) & neutral current + QCD & 5 & 214{,}372 & project engine (MadGraph-endorsed) \\
    \texttt{zjet\_k3} & \(u\bar u\to Z+3g\) & neutral current + QCD & 6 & 4{,}349{,}208 & project engine (MadGraph-endorsed) \\
    \texttt{zjet\_k4} & \(u\bar u\to Z+4g\) & neutral current + QCD & 7 & 109{,}819{,}152 & project engine (MadGraph-endorsed) \\
    \multicolumn{6}{l}{\emph{Fixed-multiplicity processes}}\\
    \texttt{ttx} & \(gg\to t\bar t\) & massive QCD & 4 & 23{,}332 & MadGraph, zero width (\(10^{-14}\)) \\
    \texttt{uu\_ag} & \(u\bar u\to\gamma g\) & QED + QCD & 4 & 10{,}289 & MadGraph (\(10^{-7}\)) \\
    \texttt{ud\_Wg} & \(u\bar d\to W^+g\) & charged current & 4 & 16{,}545 & MadGraph (\(6\times10^{-8}\)) \\
    \texttt{bb\_H} & \(b\bar b\to H\) & Yukawa & 3 & 249 & project engine BBH (exact) \\
    \texttt{bb\_Zg} & \(b\bar b\to Zg\) & down-type neutral current & 4 & 16{,}931 & project engine QQZG (\(10^{-16}\)) \\
    \texttt{uu\_ddx} & \(u\bar u\to d\bar d\) & two quark lines & 4 & 5{,}414 & MadGraph (\(7\times10^{-16}\)) \\
    \texttt{ee\_ZZ} & \(e^+e^-\to ZZ\) & leptonic diboson & 4 & 23{,}049 & project engine ee\_zz (\(10^{-16}\)) \\
    \texttt{uu\_WW} & \(u\bar u\to W^+W^-\) & diboson, triple gauge & 4 & 36{,}799 & adaptation + MadGraph (\(10^{-15}/10^{-7}\)) \\
    \texttt{uu\_ZH} & \(u\bar u\to ZH\) & associated \(VH\) & 4 & 3{,}546 & project engine UUZH (\(10^{-16}\)) \\
    \texttt{WW\_ZZ} & \(W^+W^-\to ZZ\) & quartic gauge & 4 & 102{,}687 & project engine WWZZ (\(10^{-14}\)) \\
    \hline
  \end{tabular}
\caption{The twenty processes in the tree-complexity search. \(n\) is the number of external legs and the baseline is the exactly counted real-operation total of the seed; the suite total is \(195{,}010{,}588\). Each process was validated either directly against MadGraph or against a frozen project engine previously compared point by point with MadGraph. The parenthesised values give observed relative agreement. Couplings and colour matrices are frozen in the evaluator.}
  \label{tab:processes}
\end{table}

\subsection{Sensitivity to arithmetic cost weights}
\label{subsec:operation-weighting}

The selection objective assigns unit cost to each real addition, subtraction, multiplication, division, and unary negation. To test whether the result depends on that convention, we re-instrumented the frozen seed and champion and defined
\begin{equation}
C_{w}(j)=w_+N_+(j)+w_-N_-(j)+w_\times N_\times(j)+w_{\div}N_{\div}(j)+w_{\rm neg}N_{\rm neg}(j).
\label{eq:weighted-operation-cost}
\end{equation}
Across the suite the seed contains \((70{,}088{,}824,16{,}853{,}947,104{,}689{,}102,2{,}384{,}628,994{,}087)\) operations of these five types, whereas the champion contains \((1{,}493{,}791,475{,}505,2{,}014{,}228,837,103{,}571)\). In particular, the reduction in divisions is \(2849\times\), so cost models that penalise division strengthen rather than weaken the aggregate result.

We evaluated the six representative choices in Table~\ref{tab:operation-weighting} and, separately, 1000 fixed-seed log-uniform draws with \(w_+=w_-=1\), \(w_\times\in[1,4]\), \(w_{\div}\in[1,32]\), and \(w_{\rm neg}\in[0,1]\). The named models give summed-suite improvements of \(47.7\)--\(57.2\times\) and geometric means of \(11.7\)--\(12.4\times\). Across the random ensemble the respective medians are \(51.3\times\) and \(12.0\times\), and none of the 1000 draws makes any individual process worse. The smallest process-level improvement over all named and random models is \(1.14\times\). These tests establish robustness within the stated family of arithmetic weights; they do not turn an operation count into a hardware-independent runtime model.

\begin{table}[H]
\centering
\scriptsize
\setlength{\tabcolsep}{3.5pt}
\begin{tabular}{lccccc|cc}
\toprule
Cost model & \(w_+\) & \(w_-\) & \(w_\times\) & \(w_\div\) & \(w_{\rm neg}\) & Suite reduction & Geometric mean \\
\midrule
Equal weights & 1 & 1 & 1 & 1 & 1 & \(47.7\times\) & \(11.7\times\) \\
Free negation & 1 & 1 & 1 & 1 & 0 & \(48.7\times\) & \(11.8\times\) \\
Moderate division & 1 & 1 & 1 & 4 & 0.25 & \(50.2\times\) & \(11.9\times\) \\
Expensive division & 1 & 1 & 1 & 8 & 0.25 & \(52.5\times\) & \(12.1\times\) \\
Very expensive division & 1 & 1 & 1 & 16 & 0.25 & \(57.2\times\) & \(12.4\times\) \\
Weighted multiplication & 1 & 1 & 2 & 8 & 0.25 & \(52.3\times\) & \(12.1\times\) \\
\bottomrule
\end{tabular}
\caption{Sensitivity of the tree-complexity result to six representative arithmetic cost conventions. The weights multiply the exactly counted real additions, subtractions, multiplications, divisions and unary negations in Equation~\eqref{eq:weighted-operation-cost}. Improvements are seed cost divided by champion cost; no process becomes worse under any listed model. The equal-weight row is the selection convention used in the reported search.}
\label{tab:operation-weighting}
\end{table}
\clearpage

\section{Kinematic robustness and phase-space independence}
\label{app:robustness}

A natural concern for evaluator-guided search is \emph{kinematic
overfitting}: a candidate might perform unusually well at the particular
phase-space points used by its evaluator rather than being a generally
efficient representation. This appendix separates three questions: whether
the execution route changes with continuous kinematics, whether the
performance objective transfers to unseen points, and whether the lineage
ranking survives. We answer them with fresh-kinematics audits of the frozen
champions reported in Section~\ref{sec:results}. All audit seeds are disjoint
from those used during search, scoring and the earlier checks.

\paragraph{What is selected.}
The search selects a frozen program, not a method per phase-space point. The
scientifically relevant algorithmic route is not selected point by point. The
scaling champion dispatches between its specialised and fallback algorithms
using only the discrete helicity pattern. A separate numerical safeguard may
change the unphysical gauge-reference spinor near degeneracies, but this does
not change the amplitude algorithm or its result. The tree-complexity
evaluator rejects data-dependent branching outright, so its arithmetic graph
cannot depend on the point, and the structural champion is a fixed rational
expression. The scaling evaluator's timing repeats also use different points,
because repeat \(r\) offsets the point seed by \(r\).

\subsection{Scaling run: fresh-point timing, routes and lineage ranking}

For each of the 34 scored \((k,n)\) cells we generated ten fresh points and
timed the seed and lineage champions from generations 1, 3, 18 and 47 with
five repeats per point, taking the median for each point. The champion
dispatcher was instrumented to record its route. Every champion value was
checked against the float64 Berends--Giele reference on all 340 points and
against the 40-digit \texttt{mpmath} anchor on two points per cell. The route,
timing, lineage and correctness results are summarised in
Table~\ref{tab:robustness-scaling-summary}.

\begin{table}[H]
\centering
\footnotesize
\setlength{\tabcolsep}{4pt}
\begin{tabular}{ll}
\hline
Route label (34 cells \(\times\) 10 points) & split-helicity DP on all 340; one label per cell \\
Grid geomean speed-up per point & min 859, median 892, max 895 (scoring value: 804.7) \\
Champion fastest program & 340/340 cell--point pairs; best geomean at all 10 points \\
Worst single-cell speed-up & \(107\times\) \\
Correctness & \(\leq7.3\times10^{-8}\) vs BG; \(\leq1.4\times10^{-10}\) vs mp anchor \\
\hline
\end{tabular}
\caption{Fresh-kinematics audit of the scaling champion. The route and
lineage statements cover ten unseen phase-space points in each of the 34
scored \((k,n)\) cells. Speed-up is measured relative to the BCFW seed under
the post-hoc median timing protocol. The final row gives the largest observed
relative discrepancy against the float64 Berends--Giele (BG) reference and
the 40-digit \texttt{mpmath} anchor.}
\label{tab:robustness-scaling-summary}
\end{table}

The fresh-point speed-ups are fully consistent with, and under this post-hoc
median timing protocol somewhat larger than, the archived scoring value
804.7. Across this audit, the specialised route is unchanged and the
generation-47 champion remains the fastest tested lineage member at every
cell--point pair.

\begin{figure}[ht]
  \centering
  \includegraphics[page=3,trim=85 355 80 250,clip,width=\textwidth]{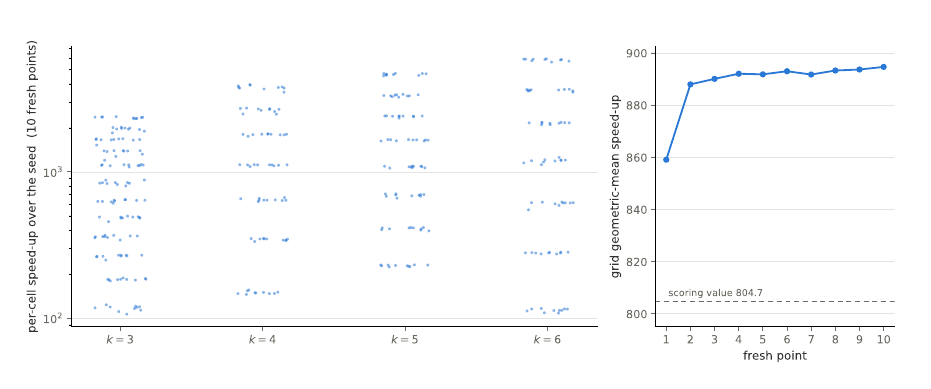}
  \caption{Kinematic robustness of the scaling champion. Left: per-cell
  speed-up over the seed at ten fresh phase-space points per scored cell,
  grouped by helicity sector. Right: the grid geometric-mean speed-up
  computed independently at each fresh point, compared with the archived
  scoring value 804.7 (dashed).}
  \label{fig:robustness-scaling}
\end{figure}

\subsection{Tree-complexity run: exact operation-count invariance}

Each of the twenty processes was run through the strict counted evaluator at
twenty fresh phase-space points, giving 400 counted evaluations per engine for
both champion and seed. We recorded the exact operation count and checked the
counted value against the independent process oracle. Table~\ref{tab:robustness-tree-summary}
summarises the results by process family.

\begin{table}[H]
\centering
\footnotesize
\setlength{\tabcolsep}{4pt}
\begin{tabular}{lcccc}
\hline
process family & processes & fresh points & distinct counts (champion/seed) & worst rel. error \\
\hline
pure gluons     & 3  & 60  & 1/1 & \(5.0\times10^{-14}\) \\
quark line      & 3  & 60  & 1/1 & \(3.8\times10^{-15}\) \\
\(Z+\)jets      & 4  & 80  & 1/1 & \(3.7\times10^{-15}\) \\
fixed processes & 10 & 200 & 1/1 & \(9.2\times10^{-14}\) \\
\hline
\end{tabular}
\caption{Fresh-kinematics audit of exact operation-count invariance in the
tree-complexity run. ``Distinct counts (champion/seed)'' gives the number of
different exact operation totals observed across the fresh points for the two
frozen programs; \(1/1\) therefore means that each program has one
point-independent count throughout that process family. The last column is
the largest relative discrepancy from the independent process oracle.}
\label{tab:robustness-tree-summary}
\end{table}

Every count equals its archived value bit for bit. The operation count is
therefore exactly constant across phase space, as required by the rejection of
data-dependent branching. The \(47.7\times\) reduction is a point-independent
statement by construction, confirmed empirically across all 400 fresh
process--point evaluations. This is the most direct audit for the
generator-relevant result: a single frozen arithmetic graph is used for each
process throughout phase space.

\subsection{Structural run: independent holdout for the cancellation metric}

The 300 scoring points per configuration were fixed during search, so the
structural objective could in principle have adapted to that sample. For each
scored configuration we therefore drew 1000 fresh points and recomputed the
experiment's own metric---the 0.95 quantile of \(\log_{10}\kappa_{\rm sum}\), including the
amplitude-zero drop rule---for seed and champion. Confidence intervals use
\(10^4\) bootstrap resamples. Table~\ref{tab:structural-holdout} compares this
holdout audit with the original frozen scoring sample. In the last column,
each cell is ``champion versus seed'' on the 300 points that remained fixed
throughout search; these are point estimates without bootstrap intervals. For
example, \(0.91\) versus \(2.12\) corresponds to 95th-percentile summation
condition numbers of approximately \(10^{0.91}\simeq8\) and
\(10^{2.12}\simeq132\), respectively. Smaller values indicate less
inter-term cancellation, but are not a direct measurement of digits lost.

\begin{table}[H]
\centering
\footnotesize
\setlength{\tabcolsep}{3pt}
\begin{tabular}{lccc}
\toprule
configuration & holdout champion \(Q_{0.95}\) [95\% CI] & holdout seed \(Q_{0.95}\) [95\% CI] & frozen sample (champion vs seed) \\
\midrule
\(n=6\), \((0,2,4)\) & 0.86 [0.79, 0.98] & 2.01 [1.86, 2.10] & 0.91 vs 2.12 \\
\(n=6\), \((0,1,3)\) & 0.51 [0.42, 0.56] & 1.68 [1.55, 1.77] & 0.40 vs 1.47 \\
\(n=7\), \((0,2,4)\) & 0.82 [0.70, 0.94] & 1.54 [1.45, 1.62] & 0.77 vs 1.58 \\
\(n=7\), \((0,1,4)\) & 0.64 [0.59, 0.68] & 1.48 [1.39, 1.53] & 0.55 vs 1.47 \\
\bottomrule
\end{tabular}
\caption{Independent holdout test of the structural cancellation metric. The holdout columns use 1000 phase-space points per helicity configuration that were not used during search; intervals are 95\% bootstrap confidence intervals. The frozen-sample column reports the champion and seed point estimates, in that order, on the original 300 fixed scoring points. The tuples give zero-based positions of the three negative-helicity legs.}
\label{tab:structural-holdout}
\end{table}

Term counts remain 2/2/3/3 at every point, and champion and seed confidence
intervals are disjoint for every configuration. Holdout and frozen-sample
values differ by at most 0.11 digits for the champion and 0.21 digits for the
seed, consistent with sampling variability in a heavy-tailed 95th percentile.
The cancellation improvement therefore persists on unseen kinematics: the
champion loses approximately 0.5--0.9 digits at the holdout 95th percentile,
compared with 1.5--2.0 for the seed.

Across these finite audits, the speed-up, execution route and lineage ranking
show no evidence of dependence on the sampled kinematics; the exact-operation
invariance is additionally guaranteed by construction.

\clearpage
\section{Controlled numerical checks of structural cancellation}
\label{app:structural-boundary-checks}

This appendix records post-search numerical checks of the structural
cancellation result. These checks were not part of the selection objective.
They compare the original CSW seed, the champion's component cells before
gluing, and the final glued champion on controlled six-point NMHV
kinematics. Their purpose is to show directly what the two figures establish,
while keeping separate the stronger statements that would require a symbolic
proof.

\paragraph{Controlled deformations.}
A one-parameter momentum-twistor deformation changes the \(\mu\)-part of one
twistor and reconstructs \(\widetilde\lambda\) through the incidence relation.
The construction preserves masslessness and momentum conservation, leaves the
angle spinors unchanged, and tunes one named denominator to
\(\lvert S\rvert=\delta\) for
\(\delta=10^{-1},\ldots,10^{-12}\).  For a term decomposition
\(A=\sum_t T_t\), the diagnostic
\[
\kappa_{\rm sum}=\frac{\sum_t |T_t|}{|\sum_t T_t|}
\]
measures the sensitivity of the final summation to independent perturbations
of its terms.

\begin{figure}[H]
  \centering
  \includegraphics[width=\linewidth]{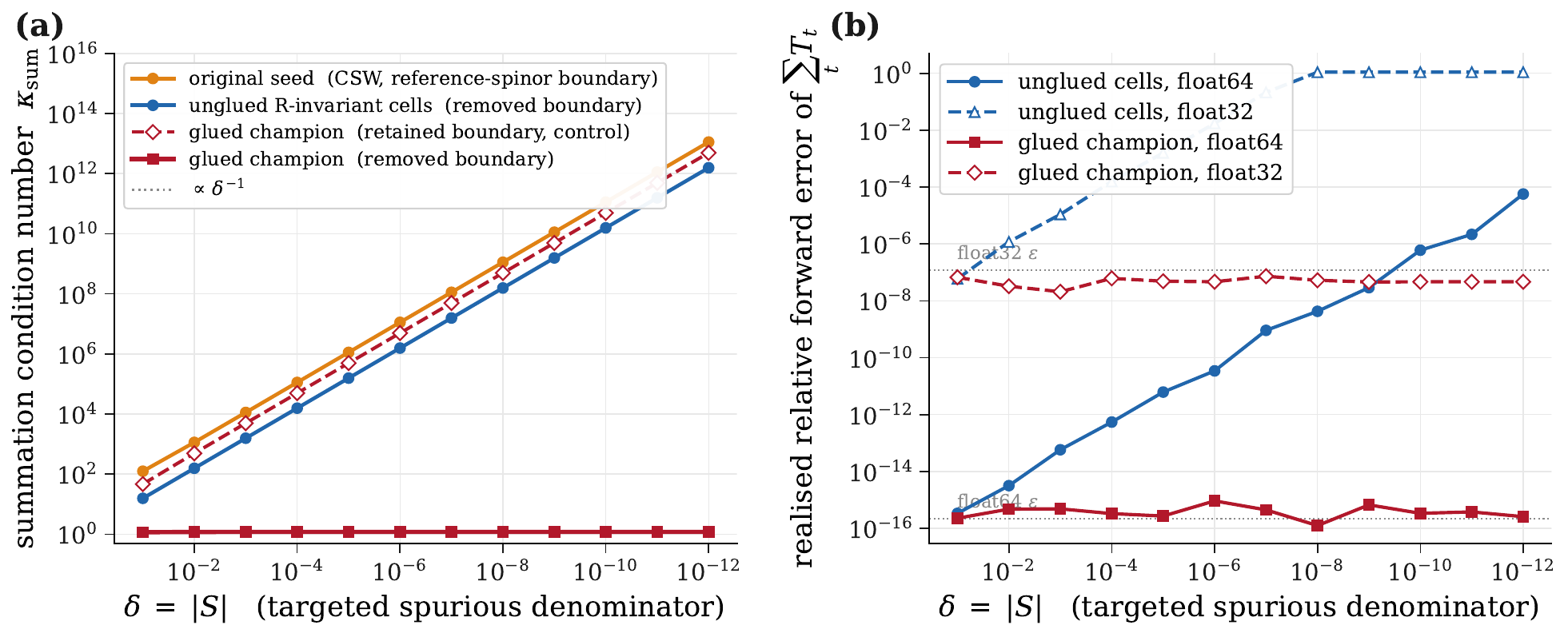}
  \caption{Controlled approaches to named denominators in the structural
  representations. Left: the CSW seed and unglued cells show
  \(\kappa_{\rm sum}\propto\delta^{-1}\) near their targeted denominators,
  whereas the champion stays \(O(1)\) at the shared boundary it removes.  The
  retained-boundary curve demonstrates residual sensitivity but is not an
  isolated one-factor test because another keyed denominator becomes small
  along that path. Right: at the removed boundary, summing the rounded
  high-precision cell terms loses float64 and float32 accuracy as
  \(\kappa_{\rm sum}\) grows, while the glued representation remains near the
  corresponding machine-precision scale. This panel diagnoses summation of
  the displayed terms; it is not a complete audit of input-level rounding in
  every implementation.}
  \label{fig:controlled-spurious-approach}
\end{figure}

Three trajectories are shown in Figure~\ref{fig:controlled-spurious-approach}:
an auxiliary-reference-spinor denominator of the CSW seed, the shared
six-point cell denominator removed by the champion, and a denominator retained
by the champion.  The removed-boundary trajectory is the cleanest test: the
targeted factor is linear in the deformation, the physical invariants and
amplitude remain bounded, and the two unglued cells develop large
opposite-sign contributions while the glued representation remains
well-conditioned.  The retained-boundary trajectory is only a qualitative
control.  On that path a second keyed denominator also becomes small---at
\(\delta=10^{-12}\) it is about \(0.24\,\delta\)---so the plot establishes
remaining sensitivity but does not isolate it to a single boundary.

At the removed boundary, the unglued condition number grows from approximately
\(15.6\) to \(1.6\times10^{12}\), while the champion remains between about
\(1.16\) and \(1.19\).  At \(\delta=10^{-12}\), the unglued rounded-term sum
has relative error about \(5.7\times10^{-5}\) in float64 and \(O(1)\) in
float32; the champion remains at the respective round-off scale.  This supports
the use of \(\kappa_{\rm sum}\) as a summation-conditioning proxy for this
trajectory.  It does not imply that the complete formula is immune to
input-level loss when small brackets are themselves formed from rounded
kinematics.

\begin{figure}[H]
  \centering
  \includegraphics[width=\linewidth]{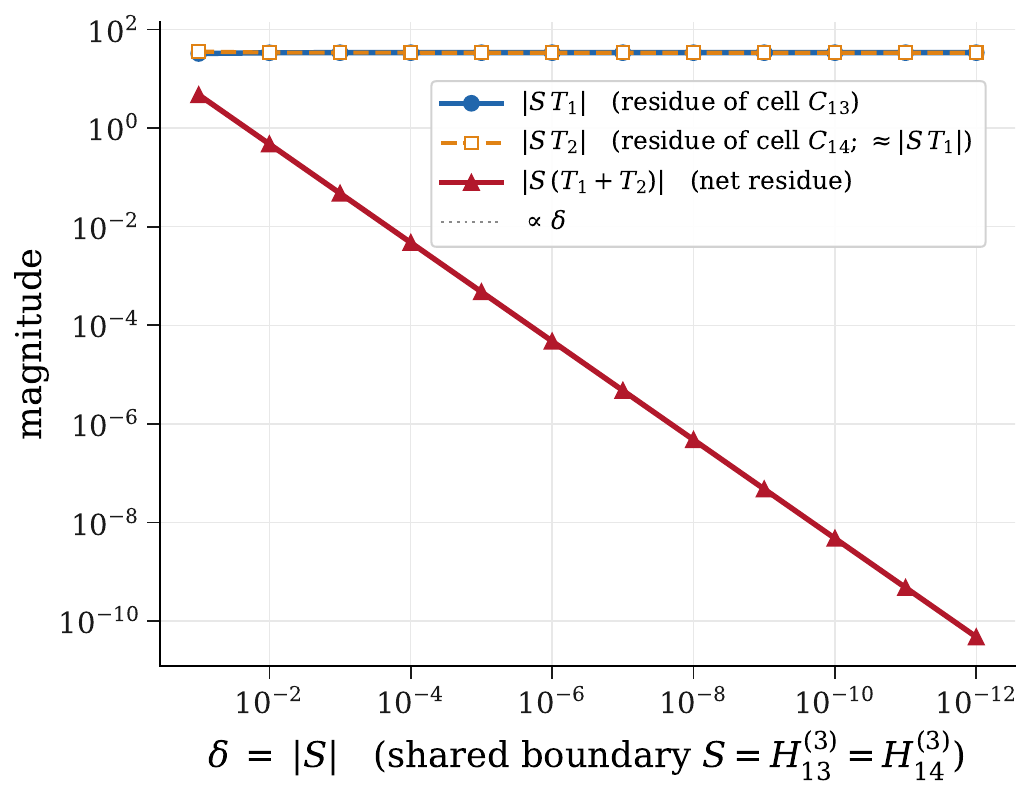}
  \caption{High-precision pointwise test of the six-point gluing operation.
  The magnitudes \(|S T_1|\) and \(|S T_2|\) approach equal nonzero values,
  while the net residue \(|S(T_1+T_2)|\) decreases linearly with \(\delta\).
  The champion's glued term agrees with \(T_1+T_2\) to order \(10^{-81}\) in
  this same-arithmetic comparison and remains finite. This is a numerical
  identity check along the displayed trajectory, not a symbolic
  divisibility proof.}
  \label{fig:gluing-residue-test}
\end{figure}

\paragraph{Pointwise gluing check.}
For the \(n=6\), negative-helicity configuration \((0,2,4)\), two neighbouring
cells \(T_1\) and \(T_2\) share a keyed denominator \(S\).  Figure
\ref{fig:gluing-residue-test} evaluates their residues
\(R_i=S T_i\) along the removed-boundary trajectory and compares their sum with
the champion's pointwise glued term.

At \(\delta=10^{-10}\), the residue ratio is
\[
\frac{R_1}{R_2}
=-0.9999999999077+1.05\times10^{-10}i,
\]
and the net residue decreases linearly with \(S\).  Repeated high-precision
checks place the relative difference between the glued term and \(T_1+T_2\)
at order \(10^{-81}\).  The released implementation nevertheless constructs
the quotient by pointwise complex division and then removes the shared factor
key; it does not contain a symbolic polynomial-division certificate.
Accordingly, the figure is strong numerical evidence for the cancellation on
this controlled family, but not a proof of divisibility over the full boundary
hypersurface.

The quoted \(10^{-81}\) agreement compares two algebraic organisations on the
same high-precision generated point.  It should not be described as an
independent 80-digit Berends--Giele validation: the released high-precision
reference wrapper converts input spinors through Python \texttt{complex}
before its \texttt{mpmath} recursion.  The ordinary correctness claims remain
those of the frozen evaluator and the independent holdout tests above.

\bibliographystyle{utphys}

\bibliography{scattering_amplitudes_program_search_arxiv}

\end{document}